\PassOptionsToPackage{numbers, compress}{natbib}
\documentclass{article}

\usepackage[preprint]{neurips_2026}

\usepackage[utf8]{inputenc}
\usepackage[T1]{fontenc}
\usepackage{hyperref}
\usepackage{url}
\usepackage{booktabs}
\usepackage{amsfonts}
\usepackage{amsmath}
\usepackage{nicefrac}
\usepackage{microtype}
\usepackage{xcolor}
\usepackage{graphicx}
\usepackage{tikz}
\usepackage{array}
\usepackage{enumitem}
\usepackage{float}
\usepackage[most]{tcolorbox}
\usepackage{tabularx}
\usetikzlibrary{positioning, arrows.meta, shapes, calc}

\hypersetup{hypertexnames=false}

\definecolor{CGRPromptBack}{HTML}{FFF7ED}
\definecolor{CGRPromptFrame}{HTML}{D08B5B}
\definecolor{CGRAuditBack}{HTML}{E8FAF3}
\definecolor{CGRAuditFrame}{HTML}{5AA37B}
\definecolor{CGRExampleBack}{HTML}{F5E8FF}
\definecolor{CGRExampleFrame}{HTML}{A66AC7}
\definecolor{boxOrange}{HTML}{FFF0E0}
\definecolor{boxBlue}{HTML}{DCE6F2}
\definecolor{boxPink}{HTML}{FADEE1}
\definecolor{boxYellow}{HTML}{FCEAB8}
\definecolor{boxGreen}{HTML}{D8EEDC}
\definecolor{boxTeal}{HTML}{D9F0ED}

\newtcolorbox{promptbox}[1]{
  colback=CGRPromptBack,
  colframe=CGRPromptFrame,
  title={#1},
  fonttitle=\bfseries,
  coltitle=black,
  boxrule=0.6pt,
  arc=1mm,
  left=1mm,
  right=1mm,
  top=1mm,
  bottom=1mm
}

\newtcolorbox{auditbox}[1]{
  colback=CGRAuditBack,
  colframe=CGRAuditFrame,
  title={#1},
  fonttitle=\bfseries,
  coltitle=black,
  boxrule=0.6pt,
  arc=1mm,
  left=1mm,
  right=1mm,
  top=1mm,
  bottom=1mm
}

\newtcolorbox{examplecard}[1]{
  colback=CGRExampleBack,
  colframe=CGRExampleFrame,
  title={#1},
  fonttitle=\bfseries\small,
  coltitle=black,
  boxrule=0.6pt,
  arc=1mm,
  left=1mm,
  right=1mm,
  top=1mm,
  bottom=1mm
}

\newcommand{\metricstd}[2]{#1{\scriptsize\,\(\pm #2\)}}
\title{Code-Guided Reasoning for Small Language Models: Evaluating Executable MCQA Scaffolds}

\author{%
 Prateek Biswas\\
  IBM\\
  New York City,NY\\
  \texttt{prateek.biswas@ibm.com} \\
  \And
  Dhaval Patel\\
  IBM\\
  Yorktown Height,NY\\
  \texttt{pateldha@us.ibm.com} \\
  \And
  Vedant Khandelwal\\
  University of South Carolina\\
  Columbia, SC\\
  \texttt{vedant@email.sc.edu} \\
  \And
  Amit Sheth\\
  University of South Carolina\\
  Columbia, SC\\
  \texttt{amit@sc.edu} \\
}

\author{%
  Prateek Biswas$^{1}$\thanks{ Corresponding author  \texttt{prateek.biswas@ibm.com}} \quad 
  Dhaval Patel$^{1}$ \quad
  Vedant Khandelwal$^{2}$ \quad
  Shuxin Lin$^{1}$ \quad
  Amit Sheth$^{2}$\\[1pt]
  $^{1}$IBM  ~ ~ $^{2}$Artificial Intelligence Institute at University of South Carolina \\
}

\begin{document}

\maketitle

\begin{abstract}

Multiple-choice QA benchmarks usually evaluate small language models (SLMs) as direct answerers, but deployed language-model systems increasingly rely on external scaffolds such as tools, code, and repeated model calls. We introduce Code- Guided Reasoning (CGR), an evaluation protocol and generated-program resource for measuring when executable reasoning scaffolds improve SLM performance on MCQA tasks. CGR standardizes six components: a normalized item interface, a direct solver prompt, a generator prompt, a Python scaffold, solver-call and extraction helpers, and a three-channel result record. On 20,498 retained result rows from a locally prepared MCQA bundle and six metadata-registered solver models, the observed non-zero-baseline partition shows 66.21\% macro assisted accuracy versus 38.11\% direct accuracy, a +28.10 percentage-point difference with a pair-bootstrap interval of [20.32, 36.43]. Under a stricter $A_b>30\%$ direct-signal gate, the macro difference is +14.11 points. These estimates are descriptive. Assisted inference uses a larger solver-call budget, answer extraction is brittle, Time-MQA contains the observed regressions, and some generated programs violate the no-hard-coding instruction. CGR provides the trace package needed to interpret these results, including direct, assisted, and generator-side answers, partition definitions, generated programs, response metadata, and audits.
\end{abstract}

\vspace{-0.35em}
\begin{figure}[H]
\centering
\includegraphics[width=0.94\linewidth]{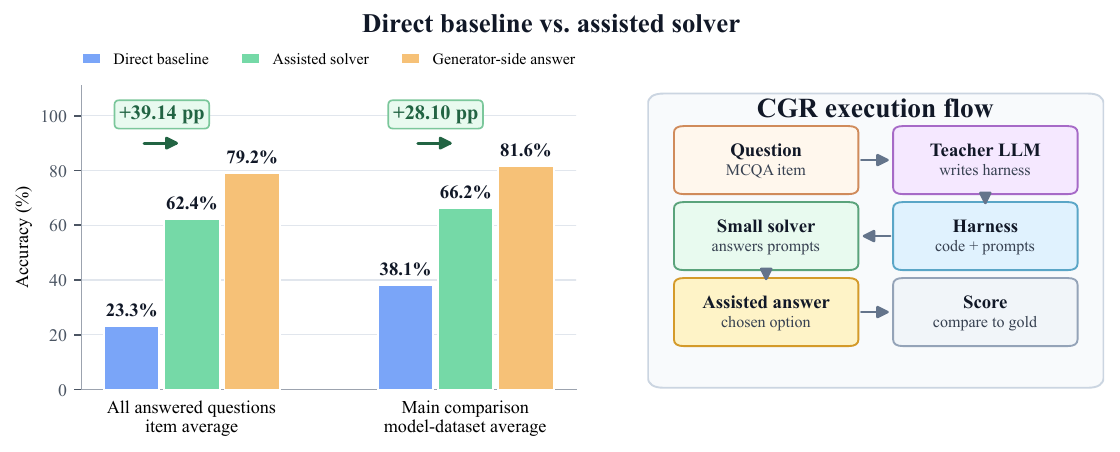}
\vspace{-0.7em}
\caption{Takeaway: CGR reports higher assisted accuracy than direct answering under the retained protocol. Left: every answered question (+39.14 pp). Right: the main comparison, averaging each dataset--solver pair once after excluding zero-direct-correct pairs (+28.10 pp). The flow panel shows the generated scaffold calling the small solver and scoring the chosen answer.}
\label{fig:first-page-teaser}
\end{figure}

\section{Introduction}

Small language models are often used for reasons that do not appear in a benchmark table. They can be cheaper to run, easier to host locally, and more practical when data or latency constraints rule out a large remote model. These systems rarely use a bare answer prompt alone. A controller may break the question into parts, call the model several times, run small computations, and then choose an option. Direct MCQA accuracy remains useful, but it does not measure this scaffolded condition.

CGR studies this interface shift. In a direct prompt, the solver emits one option letter. In CGR, a generated Python scaffold sits between the item and the solver. The scaffold can store variables, branch, call the solver helper, extract letters, and use a tiebreaker. Code-action work motivates this distinction because executable code provides control flow and inspectable state rather than a fixed text or JSON action \citep{wang2024codeact}. The evaluation question is concrete: does the same small solver behave differently when moved into this executable action space, and can the difference be audited?

The motivation is not that code alone makes a model reliable. It is that current LLMs can act as code generators and as domain-language reasoners, and many deployed workflows already combine those roles. Doctor-oriented medical-LLM work makes a related assistive distinction: the useful target is often collaboration with domain experts rather than replacing them \citep{xie2024llmsdoctors}. CGR turns that assistive premise into a narrower MCQA measurement condition: a generator writes item-specific Python that encodes domain decomposition and solver calls, while the target solver remains the model being evaluated. The resulting question is how a solver behaves when it is asked to answer through a generated domain scaffold rather than through a single natural-language option prompt.

Prior prompting and program-aided methods show that measured reasoning accuracy depends on the inference procedure \citep{wei2022chain,wang2023selfconsistency,chen2023program,gao2023pal}. Tool-use and interactive-code benchmarks add a second constraint: external calls, execution rules, and failure modes must be part of the evaluation record \citep{zhuang2023toolqa,yang2023intercode}. CGR applies that constraint to MCQA scaffolds. It does not build a reusable template memory or train a new model. It measures generated executable scaffold's as artifacts and keeps their traces visible.

For each MCQA item, a generator writes a Python function with a fixed return contract. The same target solver also answers the item directly. CGR stores the direct solver answer, the assisted solver answer produced through the generated scaffold, and the generator-side answer selected inside the scaffold. These three channels are scored separately. That separation matters because a high assisted score can come from useful decomposition, answer-format repair, extra calls, or generator-side knowledge.

The main validity choice is the non-zero-baseline partition. If a solver has no correct direct answers on a dataset, a large assisted score is hard to interpret. It may show scaffolding leverage, but it may also expose prompt-format failure or option-extraction mismatch. We therefore make the primary comparison over dataset--model pairs with at least one correct direct answer. Zero-baseline rows are reported as diagnostics, not as deployment evidence.

This paper makes three contributions:
\begin{itemize}
  \item We study executable MCQA scaffolding as an evaluation setting: the same target solver is observed under a direct option-selection prompt and inside a generated Python scaffold, with direct, assisted, and generator-side answers scored separately.
  \item We add the CGR trace package for a locally normalized MCQA bundle, recording 20,498 retained result rows across nine dataset configurations, six solver labels, generated programs, answer channels, response metadata, and source-provenance fields.
  \item We report a quantitative scaffolded-evaluation result: the observed non-zero-baseline partition improves from 38.11\% direct macro accuracy to 66.21\% assisted macro accuracy, while audits expose the larger call budget, extraction failures, answer-channel non-nesting, literal-answer patterns, and Time-MQA regressions that bound the claim.
\end{itemize}

\section{Background and Related Work}
\label{sec:background}

CGR sits between prompting, program-aided reasoning, and tool-use evaluation. Chain-of-thought prompting changes the reasoning trace requested from the model, while self-consistency changes the decoding/selection procedure \citep{wei2022chain,wang2023selfconsistency}. These methods imply a simple evaluation principle: the inference procedure belongs in the system description.

Program-aided methods make that distinction more concrete. Program of Thoughts and PAL use generated code to offload computation to an interpreter \citep{chen2023program,gao2023pal}. CodeAct frames executable code as an action format with control flow and inspectable state \citep{wang2024codeact}. CGR also uses executable code, but the code is a generated controller that can query a separate solver model, aggregate answers, and return both solver-side and generator-side judgments. This two-model structure makes it possible to compare a target SLM's direct answer with the same SLM inside a generated scaffold.

Reasoning-scaffold methods that store, retrieve, or scale thought templates show another way inference-time structure can change model behavior \citep{yang2024buffer,yang2025reasonflux}. CGR does not maintain a template memory, train a navigator, or optimize template trajectories; it evaluates freshly generated executable scaffolds as an auditable MCQA condition.

Tool-use and interactive-code benchmarks emphasize that observed language-model performance depends on external operations, tool APIs, and execution assumptions \citep{zhuang2023toolqa,yang2023intercode}. CGR inherits those concerns. A high assisted score may reflect useful decomposition, but it may also reflect a fragile answer extractor, extra inference budget, or generated code that leaks the answer. The evaluation therefore must report the scaffold contract and its violations, not just aggregate accuracy.

Benchmark papers also show why provenance has to be explicit: dataset origins, construction and conversion choices, option formats, and failure modes determine how readers interpret scores \citep{wang2024mmlupro,mihaylov2018openbookqa,kong2025timemqa}. Recent frontier reasoning benchmarks such as Humanity's Last Exam adopt a related design pattern, emphasizing hard, closed-ended, auditable tasks with clear scoring rules \citep{phan2025hle}. CGR asks an orthogonal evaluation question: for a fixed dataset--solver pair, how does the same solver behave when moved from a direct prompt into an inspectable executable scaffold?

Several evaluation resources isolate a construct that aggregate benchmarks miss and report the failures that delimit its interpretation. APPS uses executable tests for code generation \citep{hendrycks2021apps}; InterCode adds interactive execution feedback \citep{yang2023intercode}; DataComp fixes model and training choices to study dataset design \citep{gadre2023datacomp}; and DecodingTrust treats trustworthiness as an audit suite rather than a single score \citep{wang2023decodingtrust}. CGR similarly contributes a measurement setting for executable assistance, not a new solver model.

CGR is therefore an evaluation protocol rather than a model architecture claim. It asks whether a given solver, on a given dataset, changes behavior when embedded in generated executable scaffolds. The answer depends on dataset domain, solver model, prompt format, extraction rules, and generated-program quality. Outputs, generated code, dataset provenance, and partition definitions are part of the reported evidence.

\section{Datasets}
\label{sec:datasets}

The retained experiments use a locally prepared normalized MCQA bundle. Each source item has a common item id, question text, option list, option ids, and correctness flags. We treat the local records, source benchmark papers or official pages, generated programs, and retained execution traces as evidence. CureBenchPhase2QA appears in experiment metadata, but no retained solver results from that configuration enter the final analysis.

Table~\ref{tab:datasets} reports the evaluated configurations. It separates source item counts from retained registered result rows because solver coverage differs across datasets and models. The provenance column records whether each configuration is tied to a source paper, official page, or author-prepared local subset.

\begin{table*}[t]
\centering
\caption{Evaluated local MCQA configurations and provenance evidence. Source items are local configuration sizes from metadata; retained rows are registered solver-result rows.}
\label{tab:datasets}
\small
\resizebox{\textwidth}{!}{
\begin{tabular}{llrrrl}
\toprule
ID & Local config & Year & Source items & Retained rows & Provenance source \\
\midrule
\texttt{aime} & \texttt{aime2025QA} & 2025 & 30 & 173 & AIME I/II official pages \citep{aops2025aimei,aops2025aimeii} \\
\texttt{medQA} & \texttt{medQA} & 2021 & 500 & 2,957 & MedQA paper \citep{jin2021medqa} \\
\texttt{phyQA} & \texttt{physicsQA} & 2024 & 45 & 270 & PhysicsQA/MoRA paper \citep{jaiswal2024physicsqa} \\
\texttt{MMLUPro} & \texttt{MMLUPro500} & 2024 & 500 & 2,943 & MMLU-Pro paper \citep{wang2024mmlupro} \\
\texttt{SGPQA} & \texttt{SuperGPQA} & 2025 & 500 & 2,955 & SuperGPQA paper \citep{map2025supergpqa} \\
\texttt{TMQA} & \texttt{TimeMQA} & 2025 & 500 & 2,952 & Time-MQA paper \citep{kong2025timemqa} \\
\texttt{CBQA} & \texttt{CorrectBenchQA} & 2026 & 494 & 2,259 & CorrectBench Paper \cite{tie2025correctbench} \\
\texttt{OBQA} & \texttt{OpenBookQA} & 2018 & 500 & 2,991 & OpenBookQA paper \citep{mihaylov2018openbookqa} \\
\texttt{FSIQ\_RL} & \texttt{FailureSensorIQ} & 2025 & 500 & 2,998 & FailureSensorIQ paper \citep{constantinides2026failuresensoriq} \\
\bottomrule
\end{tabular}}
\end{table*}

The datasets cover different reasoning regimes. MMLU-Pro is a harder, reasoning-focused variant of MMLU with expanded answer choices and expert review \citep{wang2024mmlupro}. OpenBookQA tests application of elementary science facts plus common knowledge \citep{mihaylov2018openbookqa}. SuperGPQA targets graduate-level knowledge across many disciplines \citep{map2025supergpqa}. Time-MQA frames time series analysis as natural-language question answering over temporal data \citep{kong2025timemqa}. MedQA consists of medical board-style questions \citep{jin2021medqa}, while PhysicsQA is the physics dataset used in a refinement-agent study \citep{jaiswal2024physicsqa}. The AIME configuration wraps 30 contest problems from 2025 AIME I and II; the source pages state that the problems are copyrighted by the Mathematical Association of America, so we do not reproduce problem text here \citep{aops2025aimei,aops2025aimeii}.

FailureSensorIQ requires separate scope because the benchmark targets Industry 4.0 reasoning over failure modes, sensor data, and relationships across industrial assets \citep{constantinides2026failuresensoriq}. We label \texttt{FSIQ\_RL} as industrial sensor analytics and root-cause reasoning. The domain is hallucination-sensitive because a generated program or solver can produce plausible but unsupported links between a symptom, a sensor, and a failure mode. CGR results on this dataset evaluate scaffolded MCQA behavior; they do not establish safety for operational diagnosis without expert validation.

CGR does not claim a new public source-question corpus. The object of study is the measurement package around those questions: direct and assisted outputs, generator prompts, generated Python scaffold's, answer extraction, response metadata, and partition definitions. The source datasets define the task content; CGR defines the executable-assistance measurement setting.

\section{Methodology}
\label{sec:methodology}

\begin{figure}[t]
\centering
\resizebox{\linewidth}{!}{%
\begin{tikzpicture}[
    >={Stealth[length=3mm, width=2mm]},
    box/.style={
        rectangle,
        rounded corners=3pt,
        draw=black,
        thick,
        align=center,
        minimum height=0.9cm,
        font=\sffamily\large,
        inner sep=6pt
    },
    mcqa/.style={box, fill=boxOrange},
    baseline/.style={box, fill=boxBlue},
    gold/.style={box, fill=boxPink, minimum width=2.8cm, minimum height=1.2cm},
    exec/.style={box, fill=boxGreen},
    gen/.style={box, fill=boxYellow},
    audit/.style={box, fill=boxTeal, minimum width=12cm, minimum height=0.7cm},
    lbl/.style={font=\sffamily\large\bfseries}
]

\node[mcqa] (mcqa) {MCQA\\item};
\node[gen, right=1.5cm of mcqa, yshift=-1.7cm] (genllm) {Generator\\LLM};
\node[exec, right=0.8cm of genllm] (ps) {Python\\scaffold};
\node[exec, right=0.8cm of ps] (executor) {Executor with\\solver calls};
\node[exec, right=0.8cm of executor] (aa) {Assisted\\answer};

\node[baseline, right=1.5cm of mcqa, yshift=1.7cm] (ds) {Direct solver\\SLM};
\node[baseline] (da) at (aa |- ds) {Direct\\answer};
\node[baseline] (ae) at ($(ds)!0.5!(da)$) {Answer\\extractor};

\node[gold, right=2.0cm of aa, yshift=1.7cm] (gl) {Gold\\comparison};

\coordinate (LeftBoundary) at ($(ds.west)+(-0.5, 0)$);
\coordinate (RightBoundary) at ($(aa.east)+(0.5, 0)$);

\draw[draw=black, rounded corners=5pt, thick]
    ($(LeftBoundary |- ds.north)+(0, 0.8)$) rectangle ($(RightBoundary |- ds.south)+(0, -0.4)$);
\node[lbl, color=blue!70!black] at ($(LeftBoundary |- ds.north)!0.5!(RightBoundary |- ds.north) + (0, 0.45)$) {Direct baseline};

\draw[draw=black, rounded corners=5pt, thick]
    ($(LeftBoundary |- genllm.north)+(0, 0.8)$) rectangle ($(RightBoundary |- aa.south)+(0, -0.4)$);
\node[lbl, color=green!60!black] at ($(LeftBoundary |- genllm.north)!0.5!(RightBoundary |- genllm.north) + (0, 0.45)$) {Executable assistance};

\node[font=\sffamily\large, align=center] (cnstr) at ($(ae.south)!0.5!(ae.south |- ps.north)$) {Solver sees scaffold prompts, not the GenLLM answer};

\node[audit] (audit) at ($(ps.south)!0.5!(executor.south) - (0, 1.4cm)$) {Audit trail: prompts, generated code, solver calls, responses, metadata};

\draw[->, thick, rounded corners] (mcqa.east) -- ++(0.5,0) |- (ds.west);
\draw[->, thick, rounded corners] (mcqa.east) -- ++(0.5,0) |- (genllm.west);
\draw[->, thick] (ds.east) -- (ae.west);
\draw[->, thick] (ae.east) -- (da.west);
\draw[->, thick] (genllm.east) -- (ps.west);
\draw[->, thick] (ps.east) -- (executor.west);
\draw[->, thick] (executor.east) -- (aa.west);
\draw[->, thick, rounded corners] (da.east) -| (gl.north);
\draw[->, thick, rounded corners] (aa.east) -| (gl.south);
\draw[->, dashed, thick] (ps.south) -- (audit.north -| ps.south);
\draw[->, dashed, thick] (executor.south) -- (audit.north -| executor.south);

\end{tikzpicture}%
}
\vspace{0.45em}
\begin{examplecard}{OpenBookQA fog item}
\small
\textbf{Question.} ``There is most likely going to be fog around:'' \texttt{A} a marsh, \texttt{B} a tundra, \texttt{C} the plains, \texttt{D} a desert. Gold is \texttt{A}.\par
\textbf{Scaffold action.} The generated program defines fog as near-ground air cooling to the dew point, then compares moisture availability, cooling mechanism, wind speed, and dew-point spread across the options.\par
\textbf{Run row.} Granite 4H Small emitted direct answer \texttt{E}, outside the listed options. The assisted path used analysis and verification prompts, with a tiebreaker if extracted answers disagree. Assisted answer \texttt{A}; generator-side answer \texttt{A}; difficulty 3.
\end{examplecard}
\caption{CGR evaluation flow with a concrete retained-row example. The same MCQA item is scored through a direct solver baseline and through an executable-assistance path. The executor returns the assisted answer, while the generator-side answer and generator-estimated difficulty are stored as separate diagnostic channels; solver calls do not receive the generator-side answer.}
\label{fig:architecture}
\end{figure}

Figure~\ref{fig:architecture} summarizes the CGR protocol. The direct path asks the target solver for one option letter. The assisted path asks a generator to write an item-specific Python scaffold whose fixed return contract is \texttt{(solverLLM\_answer, genLLM\_answer, genLLM\_difficulty)}. The first value is selected after solver calls, the second is the generator-side option stored by the scaffold, and the third is generator-estimated difficulty. Solver calls inside the program receive scaffold prompts, not the generator-side answer field. A retained OpenBookQA scaffold, for example, asks the solver for an analysis answer and a verification answer, extracts both option letters, and invokes a tiebreaker only when they disagree; Appendix~\ref{app:method-details} gives the code excerpt.

Each evaluation unit has the form $(q,d,m,g)$, where $q$ is the question, $d$ is the dataset configuration, $m$ is the target solver, and $g$ is the generator label. Let $p_{q,g}=G_g(q,d)$ be the generated program, let $S_m$ denote the solver API, and let $y^\star_q$ be the gold option. The direct path observes $y_b=S_m(q)$, while the assisted path executes
\[
(y_a,y_g,\hat{h})=p_{q,g}(q,d,S_m),
\]
where $y_a$ is the scaffold-selected solver answer, $y_g$ is the generator-side selected answer, and $\hat{h}$ is the generator-estimated difficulty. Correctness is evaluated after selection as $z_c(q,m,g)=\mathbf{1}\{y_c=y^\star_q\}$ for $c\in\{b,a,g\}$. The direct, assisted, and generator-side channels are therefore reported separately.

The generated programs are synthetic scaffolds, not assumed-correct explanations. The executor supplies two helper interfaces: \texttt{llm\_model(prompt, exp\_config)}, which stores response text and metadata, and \texttt{extract\_answer(response)}, which returns the first standalone capital letter \texttt{A} through \texttt{Z} or \texttt{X}. Programs may branch, compute intermediate quantities, query the solver multiple times, and select an answer from agreement, verification, or tiebreaking logic. A program becomes evidence only when paired with execution outputs and audits of interface compliance, literal-answer patterns, extraction failures, and response metadata.

The retained logs are the source for call-count and metadata claims. The response audit finds direct-call metadata for 20,490 of 20,498 rows and assisted-call metadata for 20,492 rows, but no joined generator code-generation metadata. Generator execution is therefore evidenced by saved generated programs and result records. Direct calls have mean/median/95th-percentile/max counts of 1.01/1/1/3, while assisted calls have 7.18/6/15/90. The prompt asks for at most ten solver calls, but the runtime does not enforce that inside Python; notebook-level reattempts can also rerun invalid outputs up to \texttt{solverLLM\_reattempt\_max\_ct=3}. The no-hard-coding rule and ten-call limit are therefore prompt instructions rather than runtime guarantees, so the analysis treats violations as audit findings. A positive assisted-vs-direct difference on a non-zero-baseline pair suggests that executable assistance changed a solver with some direct task signal; zero-baseline gains and assisted regressions are diagnostic boundary cases rather than deployment evidence.

\section{Experimental Setup}
\label{sec:setup}

The analysis filters outputs to solver names listed in the solver metadata and summarizes accuracy by dataset--solver pair. This retrospective metadata-registered filter excludes unrelated pilot labels, including unregistered CBQA outputs. Coverage is uneven, so aggregate rows summarize evaluated solver coverage rather than balanced benchmark averages.

\paragraph{Models.}
Table~\ref{tab:models} lists the six retained solver names. The roster combines four earlier local solvers with Gemma 4 E2B and Nemotron-3-Nano-4B; public Artificial Analysis pages document those newer roster entries, not CGR accuracy \citep{artificialanalysis2026gemma,artificialanalysis2026nemotron}. Provider-specific model ids remain in local metadata and logs.

\begin{table}[t]
\centering
\caption{Metadata-registered solver roster used in the final analysis. Model names are shown without provider id strings.}
\label{tab:models}
\small
\resizebox{\linewidth}{!}{
\begin{tabular}{lll}
\toprule
Solver name & Local run group & Roster role \\
\midrule
\texttt{Gemma 4 E2B} & Newer retained solver & Compact recent model for local-run coverage \\
\texttt{Nemotron-3-Nano-4B} & Newer retained solver & Compact recent model for local-run coverage \\
\texttt{Granite 4H Small} & Earlier retained solver & Earlier scaffolded-run solver \\
\texttt{Granite 8B Code} & Earlier retained solver & Earlier code-oriented solver \\
\texttt{Llama 3.2 11B} & Earlier retained solver & Earlier vision-text solver on text MCQA \\
\texttt{Mistral Small 3.1 24B} & Earlier retained solver & Earlier instruction-tuned solver \\
\bottomrule
\end{tabular}}
\end{table}

The solver roster is not a balanced architecture sweep or parameter-size study. Notebooks request deterministic calls, with a 2000-token solver cap and an 8192-token generator cap.  Main-run configuration JSONL files were not retained, so these settings are supported by notebook/code evidence and response metadata rather than by a complete immutable run manifest. Appendix~\ref{app:setup-details} gives the longer provenance and runtime details.

\paragraph{Metrics and partitions.}
For each evaluated item, we compare \texttt{solverLLM\_baseline\_ans}, \texttt{solverLLM\_assisted\_ans}, and \texttt{genLLM\_ans} with \texttt{correct\_ans}. A dataset--solver pair belongs to the primary observed non-zero-baseline partition if the solver has at least one correct direct baseline answer; otherwise it belongs to the zero-baseline diagnostic partition. We report micro accuracy over evaluated items for the all-row summary, but macro accuracy over dataset--solver pairs for the partitioned claims so that larger datasets do not dominate the primary result. Let $A_b(d,m)$ and $A_a(d,m)$ denote direct and assisted accuracy; thresholded checks use $\mathcal{C}_{\tau}=\{(d,m):A_b(d,m)>\tau\}$ and average $A_a(d,m)-A_b(d,m)$ over retained pairs. The stricter $A_b>30\%$ gate checks whether gains persist when direct answering already has a meaningful signal. Generator-gap closure is $\rho=(\mathcal{M}(A_a)-\mathcal{M}(A_b))/(\mathcal{M}(A_g)-\mathcal{M}(A_b))$ for the stated aggregation $\mathcal{M}$, and is reported only as a descriptive diagnostic because the generator-side answer comes from the generated-program.

We compute uncertainty for partition-level macro quantities with a percentile bootstrap over dataset--solver pairs. The intervals do not capture repeated-generation or repeated-rerun variation because the evaluation provides one retained result per item. Appendix~\ref{app:setup-details} gives the formal partition notation.

\paragraph{Audits reported with the evaluation.}
CGR accuracy is only interpretable alongside artifact checks. We therefore report response-metadata coverage, assisted/direct call imbalance, answer-extraction failures, generated-code literal-answer scans, and threshold sensitivity with the headline results. These checks document the assumptions under which the retained scores can be read; they do not make the artifact safe or causal. The reusable artifact is the trace package around existing MCQA datasets, not a new source-question corpus; Appendix~\ref{app:artifact-intended-use} gives the retained fields, redistribution constraints, and non-use cases.

\section{Results}
\label{sec:results}

\subsection{The Three Accuracy Notions Must Be Kept Separate}

Table~\ref{tab:headline} separates the direct solver baseline, assisted solver answer, and generator-side answer. Across all evaluated items, micro direct accuracy is 23.27\%, assisted accuracy is 62.41\%, and generator-side accuracy is 79.19\%. The primary result is descriptive, not a matched-budget causal estimate: in the observed non-zero-baseline macro partition, assisted accuracy is 66.21\% versus 38.11\% for direct answering, closing 64.7\% of the generator gap and gaining 28.10 percentage points. The direct baseline is a reference condition, not a cost-matched competitor, because the assisted path can make multiple solver calls and retry invalid extractions. Appendix~\ref{app:result-figures} gives the moved visual summaries.

The table is not a single ranking. The all-item micro row answers a workload question, the primary observed macro row measures scaffolded behavior where direct answering is not completely broken, and the zero-baseline row isolates cases where generated programs produce correct assisted outputs despite no direct successes. Mixing these rows into one headline would hide the validity problem that CGR is meant to expose.

\begin{table}[t]
\centering
\caption{Result summary. ``All'' is micro-averaged over evaluated items. The two partitions are macro-averaged over dataset--solver pairs; small $\pm$ entries report standard deviations across those pairs. Generator-side accuracy is diagnostic because it comes from the generated-program workflow.}
\label{tab:headline}
\small
\resizebox{\linewidth}{!}{
\begin{tabular}{llrrrr}
\toprule
Slice & Estimator & Direct & Assisted & Gen-side & Difference \\
\midrule
All evaluated items & Micro & 23.27\% & 62.41\% & 79.19\% & +39.14 pp \\
Observed non-zero baseline & Pair macro & \metricstd{38.11\%}{28.48} & \metricstd{66.21\%}{20.14} & \metricstd{81.58\%}{14.47} & \metricstd{+28.10 pp}{24.04} \\
Zero baseline & Pair macro & \metricstd{0.00\%}{0.00} & \metricstd{62.19\%}{16.12} & \metricstd{82.30\%}{12.13} & \metricstd{+62.19 pp}{16.12} \\
\bottomrule
\end{tabular}}
\end{table}

The observed non-zero-baseline improvement has a pair-bootstrap interval of [20.32, 36.43] percentage points. Table~\ref{tab:main-validity} reports validity checks that accompany the headline number, including stricter baseline thresholds, uncertainty, budget imbalance, extraction failures, and generated-code contract violations. We treat $A_b>0$ as the broad retained-run partition and $A_b>30\%$ as the stronger direct-signal check. The retained results support a bounded claim: CGR is associated with higher assisted accuracy under both gates, but the gain is not monotone across all datasets and solvers. Dataset-cluster and solver-cluster resampling remain positive but widen the uncertainty range because solver identity and dataset domain both affect the direct baseline.

\begin{table}[t]
\centering
\caption{Validity checks for the retained result.}
\label{tab:main-validity}
\small
\resizebox{\linewidth}{!}{
\begin{tabular}{ll}
\toprule
Check & Result and interpretation \\
\midrule
Primary partition & $A_b>0\%$: $\Delta_0=+28.10$ pp \\
Stricter baseline gate & $A_b>30\%$: $\Delta_{30}=+14.11$ pp \\
Uncertainty & Pair bootstrap CI [20.32, 36.43] pp; dataset-cluster CI [18.41, 38.57] pp \\
Directionality & Most pairs improve; all regressions are Time-MQA pairs \\
Inference budget & Assisted inference uses about seven times the direct solver-token budget \\
Extraction & Assisted \texttt{X} rate 15.67\% overall and 14.44\% in the primary partition \\
Generated-code audit & Literal answer patterns appear; removing mapped rows leaves $+28.11$ pp \\
\bottomrule
\end{tabular}}
\end{table}

The answer channels are not nested subsets of one another. Table~\ref{tab:channel-overlap-main} gives a same-row overlap diagnostic, not a repeated-run consistency score. In the primary partition, 180 rows have a correct assisted answer while the generator-side answer is wrong, and 2,217 rows have the reverse pattern. The generator-side answer is therefore useful as a calibration channel but should not be collapsed into assisted-solver accuracy.

\begin{table}[t]
\centering
\caption{Same-row answer-channel overlap. A/G means assisted-solver answer and generator answer.}
\label{tab:channel-overlap-main}
\small
\resizebox{\linewidth}{!}{
\begin{tabular}{lrrrr}
\toprule
Slice & Records & A/G answer agree & A correct/G wrong & G correct/A wrong \\
\midrule
All evaluated items & 20,498 & 15,477 (75.50\%) & 249 (1.21\%) & 3,688 (17.99\%) \\
Observed non-zero baseline & 13,256 & 10,183 (76.82\%) & 180 (1.36\%) & 2,217 (16.72\%) \\
Zero baseline & 7,242 & 5,294 (73.10\%) & 69 (0.95\%) & 1,471 (20.31\%) \\
\bottomrule
\end{tabular}}
\end{table}

The stricter-threshold result is important for interpretation. The $A_b>30\%$ sensitivity keeps only pairs where the solver already answers a meaningful fraction of questions directly; the remaining positive gain suggests that the observed effect is not solely a prompt-failure rescue. The smaller effect also shows why the headline should not be summarized as a universal 28-point improvement. The primary comparison does not isolate executable structure from extra inference: matched-budget direct self-consistency, chain-of-thought direct prompting, a repeated-call no-code controller, and a generator-only direct-answer baseline are outside the current experiment set. Appendix~\ref{app:expanded-results} gives the longer reading of the moved figures.

\begin{table}[t]
\centering
\caption{Audit outcomes for the retained registered-result artifact. Counts come from response logs, generated-code scans, and answer-extraction audits.}
\label{tab:audit-outcomes}
\small
\resizebox{\linewidth}{!}{
\begin{tabular}{ll}
\toprule
Audit & Observed outcome \\
\midrule
Response metadata join & 20,498 / 20,498 rows have some metadata; direct 20,490, assisted 20,492 \\
Generator-call metadata & 0 / 20,498 registered rows have joined generator code-generation metadata \\
Solver-call count & Direct mean/median/p95/max: 1.01/1/1/3; assisted: 7.23/6/15/90 \\
Token budget & Assisted solver tokens total 148.12M vs. 20.12M direct, a 7.36$\times$ ratio \\
Extraction failures & Assisted \texttt{X}: 3,212 / 20,498 rows; direct \texttt{X}: 91 / 20,498 rows \\
Error logs & 2,523 rows across 15 files; largest categories are call-limit, key, and value errors \\
Literal-answer regex & 43 / 3,569 generated Python files; mapped to 251 registered result records \\
Prompt call limit & Prompt asks for at most ten calls; runtime logs include assisted rows above that level \\
\bottomrule
\end{tabular}}
\end{table}

\subsection{Where Assistance Helps}

The largest broad-partition gains occur when direct baseline accuracy is low but nonzero, especially on MedQA, AIME, MMLU-Pro, and SuperGPQA. On MedQA, Llama 3.2 11B rises from 1.20\% to 84.57\%, Mistral Small 3.1 24B from 3.38\% to 78.22\%, and Granite 4H Small from 1.23\% to 52.46\%. AIME also shows large gains for Mistral Small 3.1 24B and Granite 8B Code, and Granite 8B Code improves by 47.78 points on SuperGPQA and 45.78 points on MMLU-Pro. Assistance also improves several already-capable pairs, including Gemma 4 E2B on MedQA (52.91\% to 91.58\%) and Nemotron-3-Nano-4B on MMLU-Pro (64.13\% to 86.77\%). Zero-baseline improvements, such as PhysicsQA with Mistral Small 3.1 24B at 75.56\% assisted accuracy, are kept separate because they may reflect prompt-format rescue, extraction behavior, or generator/controller strength rather than direct solver competence. Appendix~\ref{app:expanded-results} gives the moved improvement matrix and the detailed pair-level examples.

These gains suggest that the generated program is sometimes doing more than rescuing a malformed answer format. The present analysis does not label every generated strategy, so the supported claim is narrower: CGR records an inspectable scaffold, call trace, and answer-channel record for later strategy-level audits.

\subsection{Negative Cases: Time-MQA}

Time-MQA is the clearest boundary condition, as shown in Table~\ref{tab:tmqa}. The same dataset contains both large gains and regressions. Low-baseline solvers benefit: Llama 3.2 11B rises from 1.42\% to 56.68\%, Mistral Small 3.1 24B rises from 8.23\% to 49.40\%, and Granite 4H Small rises from 3.58\% to 21.89\%. Three more capable direct solvers regress under assistance.

\begin{table}[H]
\centering
\caption{Time-MQA contains all primary-partition regressions. The generator-side answer is near 60--62\% across the evaluated solvers, while assisted solver behavior depends strongly on the solver.}
\label{tab:tmqa}
\small
\begin{tabular}{lrrr}
\toprule
Solver & Direct & Assisted & Gen-side \\
\midrule
\multicolumn{4}{l}{Low direct baseline; assistance helps} \\
Llama 3.2 11B & 1.42\% & 56.68\% & 61.54\% \\
Mistral Small 3.1 24B & 8.23\% & 49.40\% & 60.64\% \\
Granite 4H Small & 3.58\% & 21.89\% & 59.37\% \\
\addlinespace
\multicolumn{4}{l}{Stronger direct baseline; assistance regresses} \\
Granite 8B Code & 31.70\% & 29.24\% & 59.51\% \\
Gemma 4 E2B & 61.65\% & 56.22\% & 61.45\% \\
Nemotron-3-Nano-4B & 62.25\% & 61.04\% & 62.05\% \\
\bottomrule
\end{tabular}
\end{table}

This pattern suggests a plausible but unproven mechanism: when a solver already handles a time-series QA item directly, decomposing the question into generated subprompts can add inconsistency or distract from the most relevant temporal signal. The current results support the regression observation, not a causal explanation. We therefore treat Time-MQA as a boundary condition for future controlled ablations.

\subsection{Zero-Baseline Cases Are Diagnostic}

The zero-baseline partition reaches 62.19\% macro assisted accuracy. Several zero-baseline pairs have high assisted scores, including OBQA with Llama 3.2 11B at 88.80\%, OBQA with Mistral Small 3.1 24B at 87.00\%, MMLU-Pro with Llama 3.2 11B at 79.03\%, AIME with Llama 3.2 11B at 76.67\%, and PhysicsQA with Mistral Small 3.1 24B at 75.56\%. These results show that generated programs can sometimes overcome direct-answer failures, but they are not primary evidence of solver deployability. As diagnostics, these cases are useful because they are ambiguous. Reporting them as a separate partition identifies where controlled ablations should focus.
A separate Humanity's Last Exam pilot is reported in Appendix~\ref{app:hle-pilot}; it remains outside the primary multi-solver claim.

\subsection{Dataset Patterns}

Generator-estimated difficulty provides another internal diagnostic. Direct baseline accuracy falls from 38.69\% at difficulty 1 to 12.84\% at difficulty 9, while assisted accuracy remains near or above 50\% at every difficulty value. Figure~\ref{fig:difficulty} annotates the curve with reference rows from the local example workbook, including OBQA, PhysicsQA, and HLE-style items. Because difficulty is assigned by the generator rather than by source-dataset experts, we treat it as a scaffold-internal signal rather than a calibrated item-difficulty scale.

\section{Limitations}
\label{sec:limitations}

The current evaluation is an audit of retained experiments, not a controlled causal study with repeated stochastic trials. The positive numbers therefore describe scaffolded-system behavior under the retained protocol, not an equal-budget improvement in the underlying solver.

The missing controls are specific. A matched-budget direct self-consistency baseline would test whether repeated solver calls alone explain the assisted gains. A chain-of-thought direct prompt would test whether natural-language deliberation recovers the same signal without executable state. A repeated-call no-code controller would separate multi-prompt aggregation from Python control flow, and a generator-only direct-answer baseline would measure how much of the scaffolded result is already present in the generator channel. These controls are outside the retained runs, so the paper reports CGR as an observed evaluation condition rather than as an isolated mechanism.

\begin{auditbox}{Claims Not Supported by This Artifact}
\small
The retained evidence does not support claims that CGR is equal-cost, universally beneficial, clinically or operationally safe, or causally isolated from generator knowledge and extra solver calls. It also does not show that the no-hard-coding instruction is enforced by runtime validation.
\end{auditbox}

Three issues bound the interpretation: the assisted condition uses more solver calls, answer extraction quality, and generated-code validity is audited not enforced. Appendix~\ref{app:validity-limits} gives the longer discussion.

\section{Conclusion}

CGR is an evaluation protocol for a system question that direct MCQA scores do not answer: what happens when the same small solver moves from direct option selection into an executable scaffold that can deliberate, call the solver repeatedly, and select a final answer? In the retained metadata-registered results, the answer depends on the partition and dataset. The primary observed non-zero-baseline partition shows 66.21\% assisted versus 38.11\% direct macro accuracy, and the stricter $A_b>30\%$ gate shows a +14.11-point gain.

The main use of CGR is not the assisted score by itself. It is the paired record that lets a reader ask which channel answered correctly, how much direct signal existed for that solver--dataset pair, whether the generated program made many solver calls, whether extraction failed, and whether the generated code followed the intended contract. That record changes how scaffolded MCQA results should be read. A high assisted score with a zero direct baseline is a prompt/scaffold diagnostic. A positive score under the $A_b>30\%$ gate is stronger evidence that assistance changed a solver with direct task signal. A Time-MQA regression is evidence that generated decomposition can also disrupt an already capable direct solver.

The same evidence defines the boundary of the claim. Assistance is not equal-budget, answer extraction is brittle, Time-MQA contains regressions, and generated programs sometimes violate the no-hard-coding instruction. CGR's evaluation value is the measurement frame: direct, assisted, and generator-side answers; non-zero and zero-baseline partitions; response and code audits; and negative cases in the reported result.

The next version of the protocol should turn the audit findings into enforced checks. That means option-set-aware extraction, runtime call limits, sandboxed execution, generated-code validators, repeated generated-program sampling, and matched-budget direct baselines. Those additions would move CGR from a retrospective trace audit toward a controlled benchmark for executable assistance. The current artifact is the first step: it makes the scaffolded condition measurable and exposes enough trace evidence to decide which controls are needed next.

\bibliographystyle{unsrt}
\bibliography{references}

\clearpage
\appendix
\raggedbottom
\section{Additional Analyses}
\label{app:overview}

The additional analyses below preserve the same claim scope as the primary result. CGR observes the same target solver under a direct option-selection prompt and under a generated Python skill that can express control flow, data flow, repeated calls, intermediate checks, and final selection. The appendix reports empirical behavior, representative scaffold patterns, and limits on interpretation.

We do not reproduce full source questions. Some source items are copyrighted or governed by upstream dataset terms, and full question text is not necessary for the empirical claim. Section~\ref{app:examples} gives one representative evaluated-item pattern from each dataset, covering task type, scaffold action, answer-channel behavior, and interpretation lesson.

\section{Artifact and Intended Use}
\label{app:artifact-intended-use}

The CGR artifact is the trace package around existing MCQA datasets, not a new source-question corpus. Each retained result record stores the dataset configuration, item id, gold option, direct solver answer, assisted solver answer, generator-side answer, solver and generator labels, reattempt count, and generator-estimated difficulty. The supplementary materials are organized around the prompt templates, generated Python programs, response metadata, answer-channel result records, audit tables, and plotting/regeneration scripts.

Redistribution is constrained by upstream datasets. The reusable CGR layer is the scaffold-generation and evaluation trace; source questions should be released only where upstream terms allow it, otherwise by pointer, identifier, or derived trace. AIME is a concrete example: the source pages identify the contest problems as copyrighted, so the paper does not reproduce full items.

The intended use is evaluation research: comparing direct MCQA answering with executable assistance while keeping answer channels, inference budget, and audit failures visible. The artifact is not intended for clinical decision-making, industrial root-cause diagnosis, or safe execution of arbitrary generated Python. A complete public release should add finalized hosting metadata, including Croissant-style responsible-data fields for source provenance, generated traces, intended use, non-use, and license constraints.

\section{Methodology Details}
\label{app:method-details}

Figure~\ref{fig:workflow-example} gives the code-level version of the OpenBookQA row summarized in the main methodology figure. The generated scaffold asks the solver twice, extracts option letters, and uses a tiebreaker when calls disagree. In the retained Granite 4H Small run, the direct answer was \texttt{E}; the assisted and generator-side answers were both \texttt{A}, matching the gold label. The excerpt shows why the generator-side answer must remain a separate diagnostic channel.

\begin{figure}[H]
\small
\begin{verbatim}
response1 = llm_model(prompt=analysis_prompt, exp_config=exp_config)
answer1 = extract_answer(response=response1)

response2 = llm_model(prompt=verification_prompt, exp_config=exp_config)
answer2 = extract_answer(response=response2)

if answer1 == answer2:
    solverLLM_answer = answer1
else:
    response3 = llm_model(prompt=tiebreaker_prompt, exp_config=exp_config)
    solverLLM_answer = extract_answer(response=response3)

genLLM_answer = "A"
return (solverLLM_answer, genLLM_answer, genLLM_difficulty)
\end{verbatim}
\caption{Excerpt from a generated OpenBookQA scaffold. The full artifact keeps the generated Python program and the answer-channel record; source question text is not reproduced here.}
\label{fig:workflow-example}
\end{figure}

The retained logs are the source for call-count and metadata claims. The response audit finds direct-call metadata for 20,490 of 20,498 rows and assisted-call metadata for 20,492 rows, but no joined generator code-generation metadata. Generator execution is therefore evidenced by saved generated programs and result records. The prompt asks for at most ten solver calls, but the runtime does not enforce that inside Python; notebook-level reattempts can also rerun invalid outputs up to \texttt{solverLLM\_reattempt\_max\_ct=3}. Direct calls have mean/median/95th-percentile/max counts of 1.01/1/1/3, while assisted calls have 7.18/6/15/90.

The no-hard-coding instruction is a design intent, not an enforcement mechanism. A static audit finds literal \texttt{solverLLM\_answer = "A"}-style patterns in some generated programs. Some may reflect deterministic computation, but they violate the strictest interpretation of the prompt contract, so we treat them as an audit target rather than a guaranteed property.

\section{Experimental Setup Details}
\label{app:setup-details}

The solver roster is not a balanced architecture sweep or parameter-size study. Notebooks request solver calls at temperature 0.0 with a 2000-token cap and request the generator label \texttt{opus\_4-6} at temperature 0.0 with an 8192-token cap. Provider enforcement is partly reconstructable: WatsonX passes these as provider parameters, while the LiteLLM/LM Studio path records them inside the message object. The \texttt{gemma4\_e2b} configuration is marked as an \texttt{lmstudio} solver, with run notes recording execution on a consumer laptop; we use this as provenance, not as a throughput benchmark.

Let $A_b(d,m)$, $A_a(d,m)$, and $A_g(d,m)$ denote direct-baseline, assisted-solver, and generator-side accuracy for dataset $d$ and solver $m$. The non-zero-baseline split is a minimum interpretability gate, not a reliability claim. Without any direct correct answer, a large assisted gain can reflect prompt-format mismatch, option-extraction mismatch, or generator/controller behavior rather than solver competence. We therefore compute thresholded partitions
\[
\mathcal{C}_{\tau}=\{(d,m): A_b(d,m)>\tau\}, \qquad
\Delta_{\tau}=\frac{1}{|\mathcal{C}_{\tau}|}\sum_{(d,m)\in\mathcal{C}_{\tau}}\left(A_a(d,m)-A_b(d,m)\right),
\]
where $\Delta_0$ is the primary macro improvement and $\Delta_{30}$ is the stricter gate reported in Table~\ref{tab:main-validity}. For a chosen aggregation operator $\mathcal{M}$, we also report generator-gap closure as
\[
\rho_{\mathcal{M}}=
\frac{\mathcal{M}(A_a)-\mathcal{M}(A_b)}
{\mathcal{M}(A_g)-\mathcal{M}(A_b)},
\]
only as a descriptive diagnostic when the denominator is positive. The generator-side answer helps with calibration but should not be treated as an independently deployed baseline because the generated-program workflow produces it.

We do not report independent-answer consistency: within-scaffold repeated calls and invalid-output retries are part of the scaffold strategy, not matched repeated direct trials. The primary comparison also does not isolate executable structure from extra inference. Four controls are outside the current experiment set: matched-budget direct self-consistency, chain-of-thought direct prompting, a repeated-call no-code controller, and a generator-only direct-answer baseline. Without those runs, CGR measures the observed scaffolded system, not the causal effect of Python syntax or control flow alone.

\section{Additional Result Figures}
\label{app:result-figures}

\begin{figure}[H]
\centering
\includegraphics[width=0.78\linewidth]{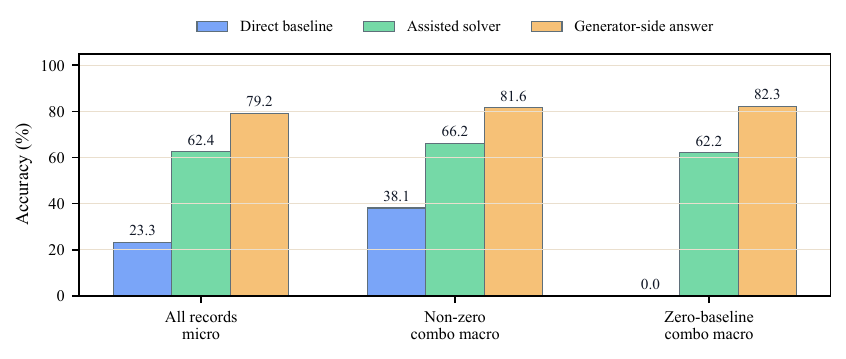}
\caption{Direct baseline, assisted solver, and generator-side answer accuracy. The x-axis ticks are the three evaluation slices: all retained records with micro averaging, observed non-zero-baseline pairs with pair-macro averaging, and zero-baseline diagnostic pairs with pair-macro averaging.}
\label{fig:performance-summary}
\end{figure}

\begin{figure}[H]
\centering
\includegraphics[width=0.98\linewidth]{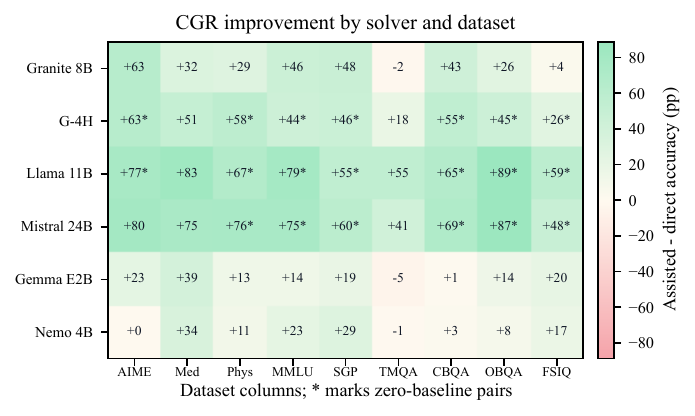}
\caption{Assisted-minus-direct accuracy by solver and dataset. Columns are datasets and rows are solvers; values are percentage-point differences. An asterisk marks zero-baseline pairs, which are diagnostic rather than primary evidence of solver capability.}
\label{fig:improvement-matrix}
\end{figure}

\begin{figure}[H]
\centering
\includegraphics[width=\linewidth]{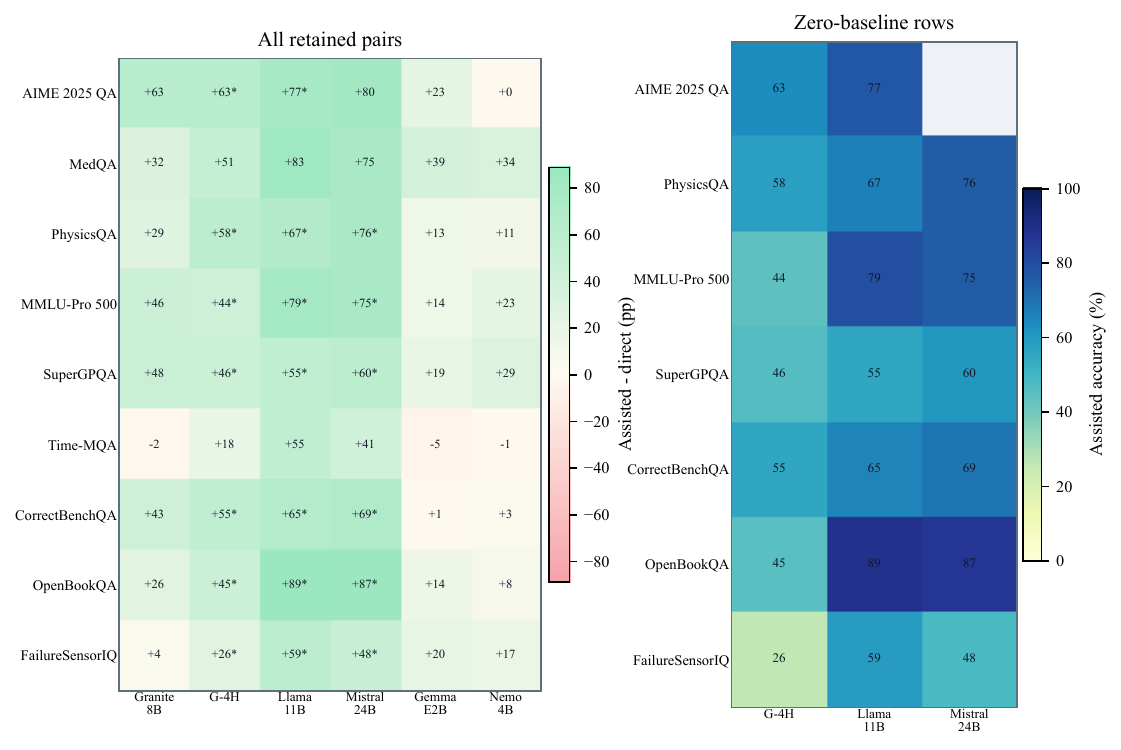}
\caption{Partitioned result matrices. The left panel reports assisted-minus-direct accuracy for all retained dataset--solver pairs; asterisks mark zero-baseline cells. The right panel reports assisted accuracy for zero-baseline diagnostic rows, where direct prompting produced no correct answers on the retained sample.}
\label{fig:core-improvement-matrix-app}
\label{fig:zero-baseline-matrix-app}
\end{figure}

\begin{figure}[H]
\centering
\includegraphics[width=0.95\linewidth]{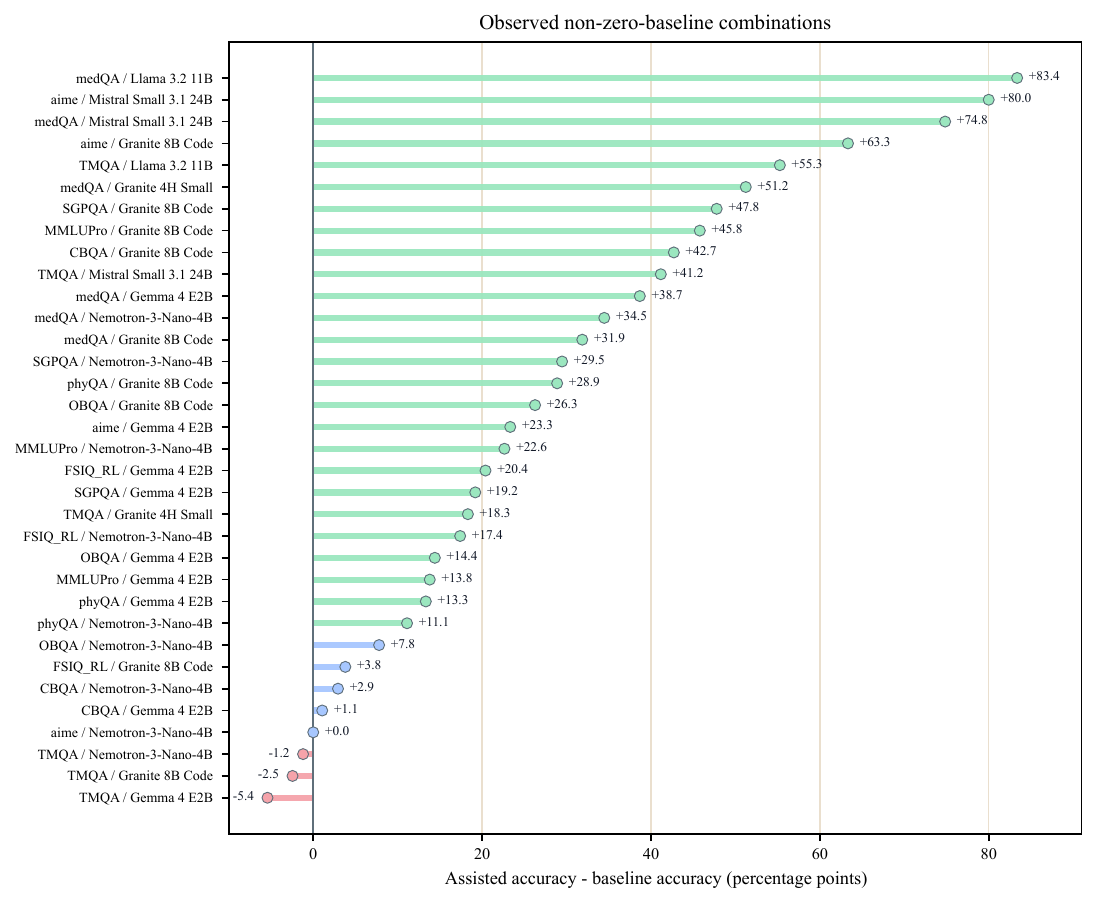}
\caption{Assisted-minus-direct accuracy for observed non-zero-baseline dataset--solver pairs. Negative bars identify the Time-MQA regressions.}
\label{fig:core-improvements}
\end{figure}

\begin{figure}[H]
\centering
\includegraphics[width=\linewidth]{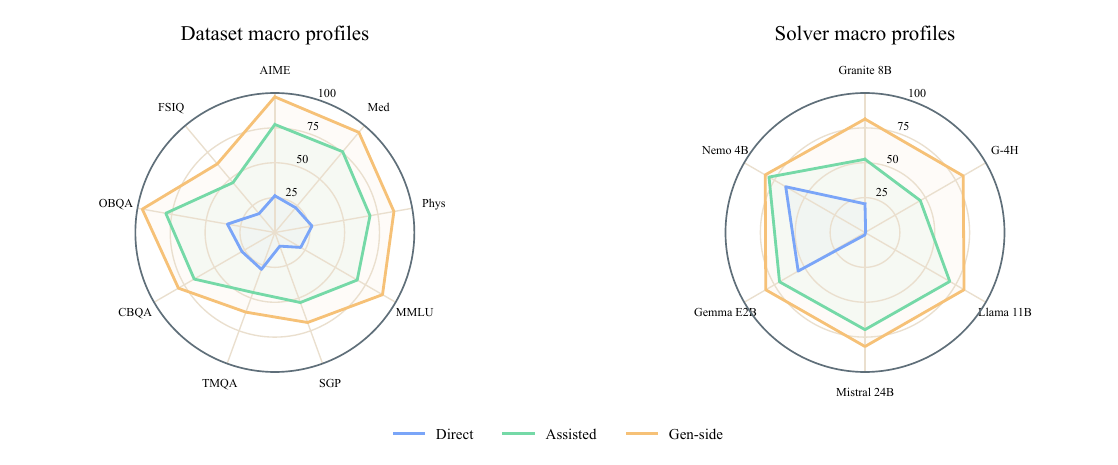}
\caption{Dataset-level and solver-level macro accuracy profiles. Each polar axis reorganizes the retained dataset--solver results: the left plot averages over solver settings within each dataset, and the right plot averages over datasets within each solver. Radius is accuracy in percent; colors match Figure~\ref{fig:performance-summary}.}
\label{fig:dataset-solver-profiles}
\end{figure}

\begin{figure}[p]
\centering
\includegraphics[height=0.86\textheight,width=\linewidth,keepaspectratio]{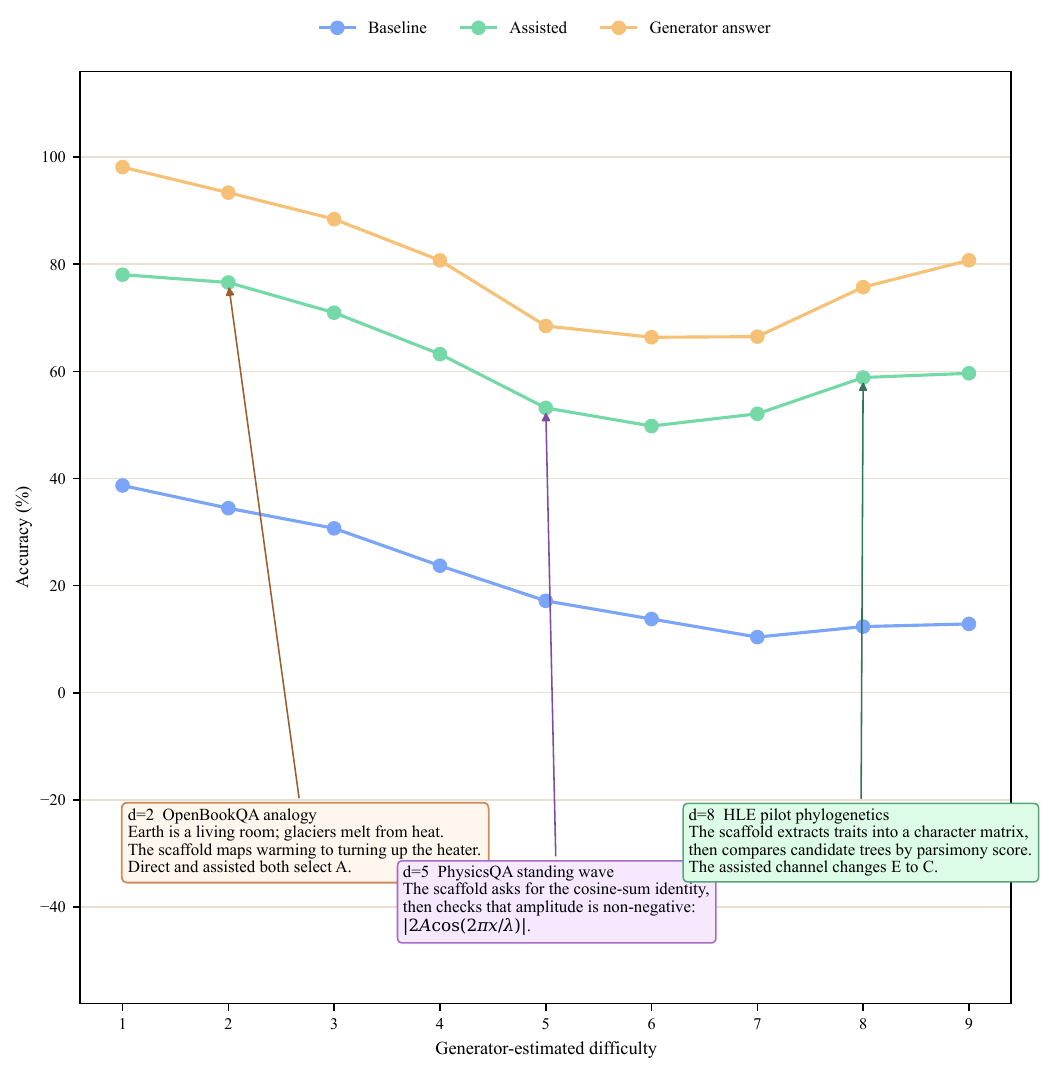}
\caption{Accuracy by generator-estimated difficulty with representative scaffold text. The examples show how a low-difficulty analogy, a mid-difficulty physics derivation, and a high-difficulty HLE pilot item are turned into structured solver prompts. Difficulty is generated metadata, not an externally validated item-difficulty annotation.}
\label{fig:difficulty}
\end{figure}
\clearpage

\begin{table}[H]
\centering
\caption{Accuracy by generator-estimated difficulty. Difficulty is assigned by the generator and functions as an internal scaffold signal rather than an independently calibrated item-difficulty label.}
\label{tab:difficulty-summary}
\small
\begin{tabular}{rrrrrr}
\toprule
Difficulty & Records & Direct & Assisted & Gen-side & Diff. \\
\midrule
1 & 747 & 38.69\% & 78.05\% & 98.13\% & +39.36 pp \\
2 & 2,772 & 34.45\% & 76.59\% & 93.36\% & +42.14 pp \\
3 & 3,917 & 30.69\% & 70.95\% & 88.41\% & +40.26 pp \\
4 & 4,657 & 23.71\% & 63.22\% & 80.72\% & +39.51 pp \\
5 & 3,387 & 17.15\% & 53.17\% & 68.47\% & +36.02 pp \\
6 & 3,270 & 13.76\% & 49.79\% & 66.36\% & +36.02 pp \\
7 & 1,396 & 10.39\% & 52.08\% & 66.48\% & +41.69 pp \\
8 & 243 & 12.35\% & 58.85\% & 75.72\% & +46.50 pp \\
9 & 109 & 12.84\% & 59.63\% & 80.73\% & +46.79 pp \\
\bottomrule
\end{tabular}
\end{table}

\section{Expanded Results Narrative}
\label{app:expanded-results}

Figure~\ref{fig:performance-summary} and Table~\ref{tab:headline} separate the direct solver baseline, assisted solver answer, and generator-side answer. Descriptively, the assisted solver closes 70.0\% of the all-item generator gap. In the observed non-zero-baseline macro partition, the assisted solver closes 64.7\% of the generator gap. The table is not a single ranking: the all-item micro row answers a workload question, the primary observed macro row measures scaffolded behavior where direct answering is not completely broken, and the zero-baseline row isolates cases where generated programs produce correct assisted outputs despite no direct successes.

Figure~\ref{fig:improvement-matrix} shows that the largest broad-partition gains occur when direct baseline accuracy is low but nonzero. MedQA has several such cases: Llama 3.2 11B rises from 1.20\% to 84.57\%, Mistral Small 3.1 24B from 3.38\% to 78.22\%, and Granite 4H Small from 1.23\% to 52.46\%. AIME also shows large gains for Mistral Small 3.1 24B and Granite 8B Code, and Granite 8B Code improves by 47.78 points on SuperGPQA and 45.78 points on MMLU-Pro.

Assistance also improves several already-capable pairs, including Gemma 4 E2B on MedQA (52.91\% to 91.58\%) and Nemotron-3-Nano-4B on MMLU-Pro (64.13\% to 86.77\%). PhysicsQA with Mistral Small 3.1 24B is different: direct accuracy is 0.00\%, while assisted accuracy reaches 75.56\%, so it belongs to the diagnostic zero-baseline reading rather than the primary solver-behavior claim. Figure~\ref{fig:dataset-solver-profiles} gives the dataset/solver profile plot; it shows the same channel separation at coarser granularity, with Time-MQA and FailureSensorIQ lower on assisted accuracy and stronger direct solvers showing smaller gains.

The stricter-threshold result is important for interpretation. The $A_b>30\%$ sensitivity keeps only pairs where the solver already answers a meaningful fraction of questions directly; the remaining positive gain suggests that the observed effect is not solely a prompt-failure rescue. The smaller effect also shows why the headline should not be summarized as a universal 28-point improvement.

\section{Humanity's Last Exam Pilot}
\label{app:hle-pilot}

Humanity's Last Exam is outside the primary multi-solver registered bundle, so it is reported as a pilot rather than folded into the headline result. The retained pilot contains one solver configuration. On 573 rows, direct accuracy is 12.57\%, assisted accuracy is 34.03\%, and the generator-side answer is also 34.03\%, for a +21.47 point assisted-minus-direct difference.

\begin{table}[H]
\centering
\caption{Humanity's Last Exam pilot result. This run is reported separately from the primary registered MCQA bundle because it contains one solver configuration rather than the multi-solver retained analysis.}
\label{tab:hle-pilot}
\small
\begin{tabular}{lrrrrr}
\toprule
Solver & Records & Direct & Assisted & Gen-side & Diff. \\
\midrule
Nemotron-3-Nano-4B & 573 & 12.57\% & 34.03\% & 34.03\% & +21.47 pp \\
\bottomrule
\end{tabular}
\end{table}

\section{Executable and Template-Based Reasoning}
\label{app:exec-template}

Prior work on executable actions and template-based reasoning shows why intermediate structure should be part of the evaluation condition. CodeAct frames executable Python as an action format whose value comes from control flow, data flow, intermediate state, tool composition, and feedback \citep{wang2024codeact}. CGR narrows those affordances to MCQA. The generated program is an executable skill around a solver SLM.

Buffer of Thoughts and ReasonFlux make a complementary point. Reasoning can be mediated by high-level structures, instantiated templates, and trajectories over simpler subproblems \citep{yang2024buffer,yang2025reasonflux}. CGR does not store a meta-buffer, retrieve thought templates, train a navigator, or optimize template trajectories. It evaluates freshly generated executable skills whose item-specific structure can parse the question, identify useful subquestions, query the solver, compare candidate answers, and select a final option.

The resulting evaluation question is direct:
\begin{quote}
Does moving the same small solver from a direct answer action into a structured executable scaffold change measured MCQA behavior, and can that change be interpreted without hiding the scaffold's failure modes?
\end{quote}

The comparison is interpretable only when three channels remain separate: the direct solver answer, assisted solver answer, and generator-side answer. A high assisted score can reflect useful decomposition. It can also reflect extra inference budget, prompt-format repair, generator-side knowledge, brittle extraction, or generated-code contract violations. CGR is evaluated as a protocol rather than as a universal improvement method.

\section{Compact Empirical Summaries}
\label{app:compact-results}

Table~\ref{tab:dataset-macro} gives a dataset-level macro summary over solver settings. MedQA and AIME have the largest average changes, while Time-MQA is mixed because it combines low-baseline improvements with regressions for stronger direct solvers.

\enlargethispage{\baselineskip}
\begin{table}[H]
\centering
\caption{Dataset-level macro accuracy over solver settings. Values are percentages; Diff. is assisted minus direct. Small $\pm$ entries are standard deviations across solver settings.}
\label{tab:dataset-macro}
\small
\setlength{\tabcolsep}{3pt}
\begin{tabular}{lrrrrrr}
\toprule
Dataset & Combos & Records & Direct & Assisted & Generator-side & Diff. \\
\midrule
AIME 2025 QA & 6 & 173 & \metricstd{26.33\%}{35.96} & \metricstd{77.44\%}{10.75} & \metricstd{97.22\%}{1.36} & \metricstd{+51.11 pp}{32.16} \\
MedQA & 6 & 2,957 & \metricstd{23.21\%}{26.20} & \metricstd{75.62\%}{17.83} & \metricstd{93.70\%}{0.55} & \metricstd{+52.41 pp}{21.89} \\
PhysicsQA & 6 & 270 & \metricstd{27.04\%}{32.79} & \metricstd{69.26\%}{10.18} & \metricstd{86.67\%}{0.00} & \metricstd{+42.22 pp}{28.04} \\
MMLU-Pro 500 & 6 & 2,943 & \metricstd{21.39\%}{32.23} & \metricstd{68.17\%}{17.64} & \metricstd{89.11\%}{0.64} & \metricstd{+46.78 pp}{26.61} \\
SuperGPQA & 6 & 2,955 & \metricstd{10.40\%}{15.28} & \metricstd{53.42\%}{7.79} & \metricstd{68.66\%}{0.22} & \metricstd{+43.01 pp}{15.69} \\
Time-MQA & 6 & 2,952 & \metricstd{28.14\%}{28.33} & \metricstd{45.75\%}{16.24} & \metricstd{60.76\%}{1.12} & \metricstd{+17.61 pp}{25.53} \\
CorrectBenchQA & 6 & 2,259 & \metricstd{27.28\%}{34.20} & \metricstd{66.76\%}{7.00} & \metricstd{79.81\%}{0.38} & \metricstd{+39.48 pp}{30.47} \\
OpenBookQA & 6 & 2,991 & \metricstd{34.51\%}{41.72} & \metricstd{79.39\%}{20.71} & \metricstd{96.49\%}{0.48} & \metricstd{+44.88 pp}{35.65} \\
FailureSensorIQ & 6 & 2,998 & \metricstd{17.63\%}{22.52} & \metricstd{46.69\%}{20.21} & \metricstd{64.18\%}{1.15} & \metricstd{+29.06 pp}{20.46} \\
\bottomrule
\end{tabular}
\end{table}

Table~\ref{tab:solver-macro} summarizes results by solver. Low direct baselines can produce large assisted gains, but already-capable solvers can also improve. The strongest direct solvers have smaller gains because their direct baselines are higher.

\begin{table}[H]
\centering
\caption{Solver-level macro summaries over retained dataset pairs. Small $\pm$ entries are standard deviations across dataset pairs for the same solver. The run group marks the local experiment grouping used in this draft and makes no public model-age or leaderboard claim.}
\label{tab:solver-macro}
\small
\resizebox{\linewidth}{!}{
\begin{tabular}{llrrrrr}
\toprule
Solver & Run group & Pairs & Records & Direct & Assisted & Diff. \\
\midrule
Nemotron-3-Nano-4B & Newer retained solver & 9 & 3,431 & \metricstd{65.62\%}{18.28} & \metricstd{79.47\%}{13.48} & \metricstd{+13.85 pp}{12.95} \\
Gemma 4 E2B & Newer retained solver & 9 & 3,443 & \metricstd{55.32\%}{16.19} & \metricstd{70.74\%}{16.23} & \metricstd{+15.42 pp}{12.70} \\
Llama 3.2 11B & Earlier retained solver & 9 & 3,437 & \metricstd{0.29\%}{0.58} & \metricstd{70.19\%}{12.47} & \metricstd{+69.90 pp}{12.50} \\
Mistral Small 3.1 24B & Earlier retained solver & 9 & 3,402 & \metricstd{1.66\%}{2.86} & \metricstd{69.64\%}{14.12} & \metricstd{+67.97 pp}{15.18} \\
Granite 8B Code & Earlier retained solver & 9 & 3,425 & \metricstd{20.53\%}{12.07} & \metricstd{52.53\%}{19.60} & \metricstd{+32.00 pp}{21.10} \\
Granite 4H Small & Earlier retained solver & 9 & 3,360 & \metricstd{0.53\%}{1.21} & \metricstd{45.77\%}{13.94} & \metricstd{+45.24 pp}{14.66} \\
\bottomrule
\end{tabular}}
\end{table}

Table~\ref{tab:best-assisted-by-dataset} reports the best observed assisted row for each dataset. This table is descriptive. It should not be read as a deployment recommendation because solver coverage, inference budget, uncertainty, and domain risk are not balanced.

\begin{table}[H]
\centering
\caption{Best observed assisted rows by dataset. This is a descriptive retained-run table, not a deployment recommendation or balanced model ranking.}
\label{tab:best-assisted-by-dataset}
\small
\resizebox{\textwidth}{!}{
\begin{tabular}{lllrrrrl}
\toprule
Dataset & Best assisted solver & Run group & Records & Direct & Assisted & Diff. & Best positive assisted row \\
\midrule
AIME 2025 QA & Nemotron-3-Nano-4B & Newer retained solver & 23 & 91.30\% & 91.30\% & +0.00 pp & Mistral Small 3.1 24B, 83.33\%, +80.00 pp \\
MedQA & Nemotron-3-Nano-4B & Newer retained solver & 499 & 57.72\% & 92.18\% & +34.47 pp & same \\
PhysicsQA & Nemotron-3-Nano-4B & Newer retained solver & 45 & 71.11\% & 82.22\% & +11.11 pp & same \\
MMLU-Pro 500 & Nemotron-3-Nano-4B & Newer retained solver & 499 & 64.13\% & 86.77\% & +22.65 pp & same \\
SuperGPQA & Nemotron-3-Nano-4B & Newer retained solver & 492 & 34.35\% & 63.82\% & +29.47 pp & same \\
Time-MQA & Nemotron-3-Nano-4B & Newer retained solver & 498 & 62.25\% & 61.04\% & -1.20 pp & Llama 3.2 11B, 56.68\%, +55.26 pp \\
CorrectBenchQA & Nemotron-3-Nano-4B & Newer retained solver & 375 & 74.13\% & 77.07\% & +2.93 pp & same \\
OpenBookQA & Nemotron-3-Nano-4B & Newer retained solver & 500 & 88.40\% & 96.20\% & +7.80 pp & same \\
FailureSensorIQ & Gemma 4 E2B & Newer retained solver & 500 & 44.40\% & 64.80\% & +20.40 pp & same \\
\bottomrule
\end{tabular}}
\end{table}

Tables~\ref{tab:full-combo-core} and~\ref{tab:full-combo-zero} give the complete registered dataset--solver breakdowns using the current registered counts.

\begin{table}[H]
\centering
\caption{Full registered dataset--solver results for the observed non-zero-baseline partition, sorted by assisted-minus-direct difference.}
\label{tab:full-combo-core}
\small
\resizebox{\textwidth}{!}{
\begin{tabular}{llrrrrr}
\toprule
Dataset & Solver & Records & Direct & Assisted & Gen-side & Diff. \\
\midrule
MedQA & Llama 3.2 11B & 499 & 1.20\% & 84.57\% & 94.19\% & +83.37 pp \\
AIME 2025 QA & Mistral Small 3.1 24B & 30 & 3.33\% & 83.33\% & 96.67\% & +80.00 pp \\
MedQA & Mistral Small 3.1 24B & 473 & 3.38\% & 78.22\% & 93.02\% & +74.84 pp \\
AIME 2025 QA & Granite 8B Code & 30 & 20.00\% & 83.33\% & 96.67\% & +63.33 pp \\
Time-MQA & Llama 3.2 11B & 494 & 1.42\% & 56.68\% & 61.54\% & +55.26 pp \\
MedQA & Granite 4H Small & 488 & 1.23\% & 52.46\% & 93.44\% & +51.23 pp \\
SuperGPQA & Granite 8B Code & 496 & 3.02\% & 50.81\% & 68.75\% & +47.78 pp \\
MMLU-Pro 500 & Granite 8B Code & 498 & 2.41\% & 48.19\% & 88.76\% & +45.78 pp \\
CorrectBenchQA & Granite 8B Code & 377 & 24.67\% & 67.37\% & 79.58\% & +42.71 pp \\
Time-MQA & Mistral Small 3.1 24B & 498 & 8.23\% & 49.40\% & 60.64\% & +41.16 pp \\
MedQA & Gemma 4 E2B & 499 & 52.91\% & 91.58\% & 94.19\% & +38.68 pp \\
MedQA & Nemotron-3-Nano-4B & 499 & 57.72\% & 92.18\% & 94.19\% & +34.47 pp \\
MedQA & Granite 8B Code & 499 & 22.85\% & 54.71\% & 93.19\% & +31.86 pp \\
SuperGPQA & Nemotron-3-Nano-4B & 492 & 34.35\% & 63.82\% & 68.70\% & +29.47 pp \\
PhysicsQA & Granite 8B Code & 45 & 28.89\% & 57.78\% & 86.67\% & +28.89 pp \\
OpenBookQA & Granite 8B Code & 491 & 37.07\% & 63.34\% & 96.13\% & +26.27 pp \\
AIME 2025 QA & Gemma 4 E2B & 30 & 43.33\% & 66.67\% & 96.67\% & +23.33 pp \\
MMLU-Pro 500 & Nemotron-3-Nano-4B & 499 & 64.13\% & 86.77\% & 89.38\% & +22.65 pp \\
FailureSensorIQ & Gemma 4 E2B & 500 & 44.40\% & 64.80\% & 65.60\% & +20.40 pp \\
SuperGPQA & Gemma 4 E2B & 495 & 25.05\% & 44.24\% & 68.69\% & +19.19 pp \\
Time-MQA & Granite 4H Small & 475 & 3.58\% & 21.89\% & 59.37\% & +18.32 pp \\
FailureSensorIQ & Nemotron-3-Nano-4B & 500 & 47.20\% & 64.60\% & 64.80\% & +17.40 pp \\
OpenBookQA & Gemma 4 E2B & 500 & 81.60\% & 96.00\% & 97.00\% & +14.40 pp \\
MMLU-Pro 500 & Gemma 4 E2B & 500 & 61.80\% & 75.60\% & 89.60\% & +13.80 pp \\
PhysicsQA & Gemma 4 E2B & 45 & 62.22\% & 75.56\% & 86.67\% & +13.33 pp \\
PhysicsQA & Nemotron-3-Nano-4B & 45 & 71.11\% & 82.22\% & 86.67\% & +11.11 pp \\
OpenBookQA & Nemotron-3-Nano-4B & 500 & 88.40\% & 96.20\% & 97.00\% & +7.80 pp \\
FailureSensorIQ & Granite 8B Code & 500 & 14.20\% & 18.00\% & 62.80\% & +3.80 pp \\
CorrectBenchQA & Nemotron-3-Nano-4B & 375 & 74.13\% & 77.07\% & 80.53\% & +2.93 pp \\
CorrectBenchQA & Gemma 4 E2B & 376 & 64.89\% & 65.96\% & 79.52\% & +1.06 pp \\
AIME 2025 QA & Nemotron-3-Nano-4B & 23 & 91.30\% & 91.30\% & 100.00\% & +0.00 pp \\
Time-MQA & Nemotron-3-Nano-4B & 498 & 62.25\% & 61.04\% & 62.05\% & -1.20 pp \\
Time-MQA & Granite 8B Code & 489 & 31.70\% & 29.24\% & 59.51\% & -2.45 pp \\
Time-MQA & Gemma 4 E2B & 498 & 61.65\% & 56.22\% & 61.45\% & -5.42 pp \\
\bottomrule
\end{tabular}}
\end{table}

\begin{table}[H]
\centering
\caption{Full registered dataset--solver results for the zero-baseline diagnostic partition, sorted by assisted accuracy.}
\label{tab:full-combo-zero}
\small
\resizebox{\textwidth}{!}{
\begin{tabular}{llrrrrr}
\toprule
Dataset & Solver & Records & Direct & Assisted & Gen-side & Diff. \\
\midrule
OpenBookQA & Llama 3.2 11B & 500 & 0.00\% & 88.80\% & 96.60\% & +88.80 pp \\
OpenBookQA & Mistral Small 3.1 24B & 500 & 0.00\% & 87.00\% & 96.40\% & +87.00 pp \\
MMLU-Pro 500 & Llama 3.2 11B & 496 & 0.00\% & 79.03\% & 89.52\% & +79.03 pp \\
AIME 2025 QA & Llama 3.2 11B & 30 & 0.00\% & 76.67\% & 96.67\% & +76.67 pp \\
PhysicsQA & Mistral Small 3.1 24B & 45 & 0.00\% & 75.56\% & 86.67\% & +75.56 pp \\
MMLU-Pro 500 & Mistral Small 3.1 24B & 493 & 0.00\% & 75.46\% & 89.45\% & +75.46 pp \\
CorrectBenchQA & Mistral Small 3.1 24B & 377 & 0.00\% & 69.50\% & 79.84\% & +69.50 pp \\
PhysicsQA & Llama 3.2 11B & 45 & 0.00\% & 66.67\% & 86.67\% & +66.67 pp \\
CorrectBenchQA & Llama 3.2 11B & 377 & 0.00\% & 65.25\% & 79.58\% & +65.25 pp \\
AIME 2025 QA & Granite 4H Small & 30 & 0.00\% & 63.33\% & 96.67\% & +63.33 pp \\
SuperGPQA & Mistral Small 3.1 24B & 487 & 0.00\% & 60.16\% & 68.99\% & +60.16 pp \\
FailureSensorIQ & Llama 3.2 11B & 500 & 0.00\% & 58.60\% & 65.00\% & +58.60 pp \\
PhysicsQA & Granite 4H Small & 45 & 0.00\% & 57.78\% & 86.67\% & +57.78 pp \\
SuperGPQA & Llama 3.2 11B & 496 & 0.00\% & 55.44\% & 68.35\% & +55.44 pp \\
CorrectBenchQA & Granite 4H Small & 377 & 0.00\% & 55.44\% & 79.84\% & +55.44 pp \\
FailureSensorIQ & Mistral Small 3.1 24B & 499 & 0.00\% & 48.10\% & 63.93\% & +48.10 pp \\
SuperGPQA & Granite 4H Small & 489 & 0.00\% & 46.01\% & 68.51\% & +46.01 pp \\
OpenBookQA & Granite 4H Small & 500 & 0.00\% & 45.00\% & 95.80\% & +45.00 pp \\
MMLU-Pro 500 & Granite 4H Small & 457 & 0.00\% & 43.98\% & 87.96\% & +43.98 pp \\
FailureSensorIQ & Granite 4H Small & 499 & 0.00\% & 26.05\% & 62.93\% & +26.05 pp \\
\bottomrule
\end{tabular}}
\end{table}

\section{Threshold, Uncertainty, and Extraction Sensitivity}
\label{app:sensitivity}

The primary analysis uses the permissive observed non-zero-baseline partition because it separates total direct-answer failures from cases with at least some direct signal. Table~\ref{tab:threshold-sensitivity-compact} shows why the result should not be summarized as a universal 28-point improvement: the gain remains positive under stricter direct-baseline gates, but it shrinks as the retained solver settings become more directly capable.

\begin{table}[H]
\centering
\caption{Sensitivity to stricter direct-baseline gates. Values are macro accuracies over retained dataset--solver pairs; small $\pm$ entries are standard deviations across retained pairs.}
\label{tab:threshold-sensitivity-compact}
\small
\begin{tabular}{rrrr}
\toprule
$A_b$ threshold & Direct & Assisted & Diff. \\
\midrule
$>0\%$ & \metricstd{38.11\%}{28.48} & \metricstd{66.21\%}{20.14} & \metricstd{+28.10 pp}{24.04} \\
$>2\%$ & \metricstd{41.67\%}{27.27} & \metricstd{66.37\%}{20.63} & \metricstd{+24.70 pp}{21.90} \\
$>5\%$ & \metricstd{49.08\%}{23.20} & \metricstd{68.27\%}{19.69} & \metricstd{+19.19 pp}{16.75} \\
$>10\%$ & \metricstd{50.71\%}{22.10} & \metricstd{69.02\%}{19.71} & \metricstd{+18.31 pp}{16.47} \\
$>20\%$ & \metricstd{53.64\%}{20.50} & \metricstd{70.62\%}{17.13} & \metricstd{+16.99 pp}{13.87} \\
$>30\%$ & \metricstd{59.59\%}{17.25} & \metricstd{73.69\%}{16.92} & \metricstd{+14.11 pp}{13.01} \\
\bottomrule
\end{tabular}
\end{table}

Table~\ref{tab:uncertainty-compact} reports uncertainty checks that broaden the resampling unit. The estimate is positive under each resampling choice, but solver-cluster uncertainty is much wider because solver identity strongly affects direct-baseline behavior.

\begin{table}[H]
\centering
\caption{Uncertainty and cluster sensitivity for the non-zero-baseline macro improvement.}
\label{tab:uncertainty-compact}
\small
\begin{tabular}{lll}
\toprule
Check & Unit & Result \\
\midrule
Primary estimate & Dataset--solver pair & +28.10 pp \\
Bootstrap 95\% interval & Dataset--solver pair & [20.32, 36.43] pp \\
Bootstrap 95\% interval & Dataset cluster & [18.41, 38.57] pp \\
Bootstrap 95\% interval & Solver cluster & [17.95, 48.47] pp \\
Leave-one range & Dataset & [22.90, 30.35] pp \\
Leave-one range & Solver & [24.50, 33.24] pp \\
\bottomrule
\end{tabular}
\end{table}

Answer extraction is a central validity issue because the extractor accepts the first standalone capital letter and returns \texttt{X} when no such letter appears. Table~\ref{tab:extraction-compact} reports extraction-failure rates rather than treating extraction as invisible plumbing. Assisted outputs fail extraction much more often than direct or generator-side outputs, which is one reason the assisted path uses reattempts.

\begin{table}[H]
\centering
\caption{Extractor failure rates. \texttt{X} means no standalone capital option letter was extracted.}
\label{tab:extraction-compact}
\small
\begin{tabular}{lrrr}
\toprule
Slice & Direct baseline & Assisted solver & Generator-side \\
\midrule
All evaluated items & 0.44\% & 15.67\% & 0.79\% \\
Observed non-zero baseline & 0.32\% & 14.44\% & 0.79\% \\
Zero baseline & 0.66\% & 17.92\% & 0.79\% \\
\bottomrule
\end{tabular}
\end{table}

\section{Artifact Audit Details}
\label{app:artifact-audit}

The main paper reports the compact audit outcomes. Table~\ref{tab:artifact-audit-summary} gives the appendix version, including response-log coverage, unavailable generator code-generation metadata, static call-site checks, literal-answer scans, and unavailable main-run config JSONL files.

\begin{table}[H]
\centering
\caption{Artifact audit summary. These checks document what is present in the retained local artifacts; they are audits, not proofs of semantic correctness or execution safety.}
\label{tab:artifact-audit-summary}
\small
\begin{tabular}{lr}
\toprule
Audit item & Value \\
\midrule
Main-run config JSONL files found & 2 \\
Response metadata files found & 19 \\
Response metadata rows & 175,374 \\
Retained rows with any response metadata & 20,498/20,498 \\
Rows with direct solver metadata & 20,490/20,498 \\
Rows with assisted solver metadata & 20,492/20,498 \\
Generator code-generation calls joined to retained rows & 0 \\
Mean direct solver calls per result & 1.01 \\
Mean assisted solver calls per result & 7.23 \\
Max assisted solver calls per result & 90 \\
Assisted/direct solver-token ratio & 7.36 \\
Generated Python files scanned & 3,569 \\
Files with >10 static \texttt{llm\_model} call sites & 2 \\
Max static \texttt{llm\_model} call sites in a file & 11 \\
Literal answer-pattern files & 43 \\
Mapped result rows using literal-pattern files & 251 \\
Core diff. after removing mapped literal-pattern rows & +28.11 pp \\
Error log rows found & 2,523 \\
\bottomrule
\end{tabular}
\end{table}

\section{Answer-Channel Overlap}
\label{app:channel-overlap}

The overlap table answers a specific diagnostic question: are assisted-solver successes a subset of generator-side successes? No. These rows are same-record channel comparisons, not repeated stochastic consistency measurements.

\begin{table}[H]
\centering
\caption{Answer-channel overlap diagnostic. These same-row overlaps are not repeated-run consistency estimates; they show where direct, assisted, and generator-side channels agree or disagree on retained records.}
\label{tab:channel-overlap}
\small
\resizebox{\linewidth}{!}{
\begin{tabular}{lrrrrrr}
\toprule
Slice & Records & A/G answer agree & A correct/G wrong & G correct/A wrong & A correct/direct wrong & Direct correct/A wrong \\
\midrule
All rows & 20,498 & 15,477 (75.50\%) & 249 (1.21\%) & 3,688 (17.99\%) & 8,895 (43.39\%) & 872 (4.25\%) \\
Observed non-zero baseline & 13,256 & 10,183 (76.82\%) & 180 (1.36\%) & 2,217 (16.72\%) & 4,521 (34.11\%) & 872 (6.58\%) \\
Zero baseline & 7,242 & 5,294 (73.10\%) & 69 (0.95\%) & 1,471 (20.31\%) & 4,374 (60.40\%) & 0 (0.00\%) \\
\bottomrule
\end{tabular}}
\end{table}

\section{Additional Validity Limits}
\label{app:validity-limits}

The current evaluation has one retained result per evaluated item, no repeated generations for uncertainty over generated programs, and no verified execution-time distribution. The bootstrap intervals quantify variability over dataset--solver pairs, not over repeated runs of the same item.

The assisted condition uses more inference than the direct baseline. Generated programs can make multiple solver calls, and the execution workflow can retry invalid extracted answers. Assisted accuracy therefore measures a scaffolded system under a larger budget, not an equal-cost replacement for direct answering. The retained artifacts do not include a matched-budget direct self-consistency or majority-vote baseline, so they do not isolate the code scaffold from the extra solver-call budget.

The answer extraction function is brittle. It returns the first standalone uppercase letter \texttt{A} through \texttt{Z}. This can incorrectly accept letters outside the available option set, and it can be triggered by incidental capital letters in noncompliant model output. Zero-baseline results are especially sensitive to this issue.

Generated-code validity is not guaranteed. The prompt prohibits hard-coding, but the literal-assignment scan found patterns consistent with direct assignment to \texttt{solverLLM\_answer}. The execution environment is an unhardened notebook-style Python environment with no proven sandbox, timeout guarantee, import allowlist, filesystem isolation, or dynamic semantic validator. Future versions should add static and dynamic validators before reporting scaffolded accuracy.

Dataset provenance is uneven. Most dataset descriptions are backed by original papers or official documentation. Local conversion and sampling choices may differ from the original datasets. High-stakes domains such as medicine and industrial root-cause analysis should be interpreted as benchmark domains only.

\section{Representative Dataset Examples}
\label{app:examples}

The examples below are representative evaluated-item patterns, not a new dataset release and not full source-question reproduction. Each card shows what the generated scaffold did and what interpretive lesson the example supports.

\begin{figure}[H]
\centering
\begin{minipage}[t]{0.32\linewidth}
\vspace{0pt}
\begin{examplecard}{AIME 2025 QA}
\small
\textbf{Task type.} Contest geometry with side ratios, reflections, and area comparison.\par
\textbf{Scaffold action.} Introduces vector coordinates, expresses reflected points algebraically, computes polygon area by cross products, and asks the solver to verify the derived option.\par
\textbf{Observed behavior.} A direct response can miss the geometric invariant, while the scaffolded path selects the computed answer.\par
\textbf{Lesson.} In this example, executable structure makes algebraic invariants explicit; the small contest slice still needs cautious interpretation.
\end{examplecard}
\end{minipage}\hfill
\begin{minipage}[t]{0.32\linewidth}
\vspace{0pt}
\begin{examplecard}{CorrectBenchQA}
\small
\textbf{Task type.} Arithmetic mixture problem with duplicate correct-looking options.\par
\textbf{Scaffold action.} Decomposes alloy ratios into component masses, simplifies the final ratio, then verifies which option label matches first.\par
\textbf{Observed behavior.} The scaffold resolves the calculation and normalizes duplicate labels to a single option.\par
\textbf{Lesson.} Some gains may reflect formatting and option-handling support alongside calculation steps.
\end{examplecard}
\end{minipage}\hfill
\begin{minipage}[t]{0.32\linewidth}
\vspace{0pt}
\begin{examplecard}{FailureSensorIQ}
\small
\textbf{Task type.} Industrial sensor question about which signal is least useful for compressor-stall detection.\par
\textbf{Scaffold action.} Builds domain context, compares pressure, vibration, temperature, and speed sensors, then asks for a final least-useful choice.\par
\textbf{Observed behavior.} The scaffold creates a structured domain rationale before answer selection.\par
\textbf{Lesson.} The scaffold organizes the sensor comparison, but plausible sensor-failure explanations can be hallucination-sensitive.
\end{examplecard}
\end{minipage}
\caption{Representative examples from AIME, CorrectBenchQA, and FailureSensorIQ.}
\label{fig:example-cards-1}
\end{figure}

\begin{figure}[H]
\centering
\begin{minipage}[t]{0.32\linewidth}
\vspace{0pt}
\begin{examplecard}{MMLU-Pro}
\small
\textbf{Task type.} History item about the shared aim of the Carlsbad Resolutions and the Concert of Europe.\par
\textbf{Scaffold action.} Supplies historical context about Metternich, separates overbroad distractors from nationalist-revolution suppression, and verifies the final option.\par
\textbf{Observed behavior.} The assisted and generator-side channels select the historically grounded option where the direct output can be invalid.\par
\textbf{Lesson.} Generated context can coincide with direct-prompt failure recovery, but the generator-side channel must remain separate.
\end{examplecard}
\end{minipage}\hfill
\begin{minipage}[t]{0.32\linewidth}
\vspace{0pt}
\begin{examplecard}{OpenBookQA}
\small
\textbf{Task type.} Elementary science item about environments where fog is likely.\par
\textbf{Scaffold action.} Frames fog formation through moisture, cooling, wind, and dew-point spread, then compares wetlands, tundra, plains, and desert options.\par
\textbf{Observed behavior.} The scaffolded answer follows the moisture-based explanation.\par
\textbf{Lesson.} The code scaffold turns this short science fact question into a structured causal check.
\end{examplecard}
\end{minipage}\hfill
\begin{minipage}[t]{0.32\linewidth}
\vspace{0pt}
\begin{examplecard}{SuperGPQA}
\small
\textbf{Task type.} Physics item about hydrogen emission transitions from an excited state.\par
\textbf{Scaffold action.} Enumerates all possible transitions, treats each atom as one cascade path, and reasons about the minimum path cover.\par
\textbf{Observed behavior.} The scaffold uses set-cover style reasoning before asking for the final option.\par
\textbf{Lesson.} In this example, the scaffold turns implicit combinatorics into an explicit search over transitions.
\end{examplecard}
\end{minipage}
\caption{Representative examples from MMLU-Pro, OpenBookQA, and SuperGPQA.}
\label{fig:example-cards-2}
\end{figure}

\begin{figure}[H]
\centering
\begin{minipage}[t]{0.32\linewidth}
\vspace{0pt}
\begin{examplecard}{Time-MQA}
\small
\textbf{Task type.} Time-series volatility classification.\par
\textbf{Scaffold action.} Computes mean, standard deviation, range, consecutive differences, and majority-votes several solver prompts.\par
\textbf{Observed behavior.} The same dataset also contains regressions for stronger direct solvers.\par
\textbf{Lesson.} Numeric preprocessing can help weak direct solvers, but generated decomposition can disturb already-capable temporal reasoning.
\end{examplecard}
\end{minipage}\hfill
\begin{minipage}[t]{0.32\linewidth}
\vspace{0pt}
\begin{examplecard}{MedQA}
\small
\textbf{Task type.} Medical board-style differential diagnosis after recent travel with fever, rash, cytopenias, and joint symptoms.\par
\textbf{Scaffold action.} Separates candidate infections, identifies distinguishing clinical features, and asks a final board-style answer prompt.\par
\textbf{Observed behavior.} The MedQA dataset-level summary reports the largest assisted-minus-direct difference.\par
\textbf{Lesson.} The result is benchmark evidence only, with no clinical-deployment support.
\end{examplecard}
\end{minipage}\hfill
\begin{minipage}[t]{0.32\linewidth}
\vspace{0pt}
\begin{examplecard}{PhysicsQA}
\small
\textbf{Task type.} Sound-wave question involving wavelength, diffraction, medium changes, interference, and Doppler shift.\par
\textbf{Scaffold action.} Computes wavelength, checks each option against physics principles, and verifies the single correct statement.\par
\textbf{Observed behavior.} The scaffold makes the quantitative scale mismatch explicit before selection.\par
\textbf{Lesson.} Executable scaffolds can combine calculation with conceptual option elimination.
\end{examplecard}
\end{minipage}
\caption{Representative examples from Time-MQA, MedQA, and PhysicsQA.}
\label{fig:example-cards-3}
\end{figure}

\begin{figure}[H]
\centering
\begin{minipage}[t]{0.62\linewidth}
\vspace{0pt}
\begin{examplecard}{Humanity's Last Exam Pilot}
\small
\textbf{Task type.} Phylogenetic parsimony over alien-species morphological traits.\par
\textbf{Scaffold action.} Extracts a binary character matrix for setose covering, claws, simple eyes, antennae, and serrate antennae, then asks the solver to compare candidate trees by minimum character-state changes.\par
\textbf{Observed behavior.} In the reference row, the direct solver selects \texttt{E}; the assisted path and generator-side channel select \texttt{C}, matching the gold label, with generator-estimated difficulty 8.\par
\textbf{Lesson.} The HLE run is useful as a hard-benchmark pilot, but it remains outside the balanced primary analysis because only one solver configuration is present.
\end{examplecard}
\end{minipage}
\caption{Representative pilot example from the HLE evaluation artifacts.}
\label{fig:hle-example-card}
\end{figure}

\section{Per-Dataset Strong-LLM Harness Examples}
\label{app:iteration-patterns}

The examples below show the generated strong-LLM harnesses behind representative rows from each dataset. Here, a harness is the generated Python scaffold for one item: it may do local computation, call the target solver through a prompt, extract answer letters, run verification or tiebreak prompts, and return the assisted solver answer, the generator-side answer, and a generator-estimated difficulty. These are illustrative reference rows, not a claim that every generated scaffold in a dataset has the same structure.

\begin{auditbox}{Harness Template}
\small
\textbf{Input.} A multiple-choice item and its answer options.\par
\textbf{Generator stage.} A strong generator LLM writes a scaffold tailored to that item.\par
\textbf{Execution stage.} The scaffold performs any local computation or structured decomposition, sends one or more prompts to the target solver, and extracts option letters.\par
\textbf{Selection stage.} The scaffold keeps an agreeing answer, verifies the strongest candidate, or asks a compact tiebreaker when calls disagree.\par
\textbf{Output.} The returned tuple records the assisted solver channel, the generator-side channel, and difficulty metadata.
\end{auditbox}

\begin{figure}[H]
\centering
\begin{minipage}[t]{0.48\linewidth}
\vspace{0pt}
\begin{examplecard}{AIME 2025 QA}
\small
\textbf{Problem shape.} Count ways to choose 8 of 16 chairs with no three consecutive occupied chairs, then report the residue modulo 1000.\par
\textbf{Strong-LLM harness.} Builds a dynamic program over chair index, selected count, and the last two occupancy bits. The computed count is converted to the answer-option residue.\par
\textbf{Iteration.} (1) run the local DP/enumeration; (2) prompt the solver with the computed count and residue; (3) extract the solver option and compare it with the generator's option map.\par
\textbf{Return.} The harness returns the solver-selected option, the DP-derived generator option, and high difficulty metadata.
\end{examplecard}
\end{minipage}\hfill
\begin{minipage}[t]{0.48\linewidth}
\vspace{0pt}
\begin{examplecard}{CorrectBenchQA}
\small
\textbf{Problem shape.} A ratio/fraction arithmetic item in which the denominator exceeds the numerator by a fixed amount.\par
\textbf{Strong-LLM harness.} Solves the algebra directly, then asks the target solver to check the relation between numerator, denominator, and option labels.\par
\textbf{Iteration.} (1) derive the variables from the ratio; (2) ask a guided solver prompt; (3) ask a verification prompt; (4) keep the agreeing answer or choose from the verified calculation.\par
\textbf{Return.} The scaffold records a low-to-mid difficulty arithmetic item where harness value comes from normalization and verification.
\end{examplecard}
\end{minipage}
\caption{Strong-LLM harness examples for AIME and CorrectBenchQA. The cards show local computation, guided solver prompting, and verification before answer extraction.}
\label{fig:harness-examples-1}
\end{figure}

\begin{figure}[H]
\centering
\begin{minipage}[t]{0.48\linewidth}
\vspace{0pt}
\begin{examplecard}{FailureSensorIQ}
\small
\textbf{Problem shape.} Identify the least useful sensor signal for compressor-stall or misalignment monitoring.\par
\textbf{Strong-LLM harness.} Creates domain context about misalignment symptoms, evaluates candidate sensor types, and asks a specialist-style verification prompt.\par
\textbf{Iteration.} (1) list relevant fault symptoms; (2) evaluate length, vibration, and leakage signals using the context; (3) ask an independent verification prompt; (4) invoke a tiebreaker if the extracted answers disagree.\par
\textbf{Return.} The harness exposes how domain decomposition can help while also making hallucination risk inspectable.
\end{examplecard}
\end{minipage}\hfill
\begin{minipage}[t]{0.48\linewidth}
\vspace{0pt}
\begin{examplecard}{MMLU-Pro}
\small
\textbf{Problem shape.} A medical-domain multiple-choice row about drug toxicity and distinguishing symptoms in an older patient.\par
\textbf{Strong-LLM harness.} Produces a clinical differential, checks medication and symptom evidence, then asks for an option-level verification.\par
\textbf{Iteration.} (1) analyze the clinical scenario; (2) focus on the drug-toxicity hypothesis; (3) compare each option against that hypothesis; (4) use a final verification or tiebreak prompt before extraction.\par
\textbf{Return.} The scaffold separates generator-side clinical reasoning from the assisted solver answer, which is important for interpreting medical-domain gains.
\end{examplecard}
\end{minipage}
\caption{Strong-LLM harness examples for FailureSensorIQ and MMLU-Pro. Both use domain-context prompts before option selection.}
\label{fig:harness-examples-2}
\end{figure}

\begin{figure}[H]
\centering
\begin{minipage}[t]{0.48\linewidth}
\vspace{0pt}
\begin{examplecard}{OpenBookQA}
\small
\textbf{Problem shape.} A short elementary-science analogy: glacier melting is mapped to a room-temperature intervention.\par
\textbf{Strong-LLM harness.} Turns the analogy into causal roles, then asks a second solver pass to view the choices as heat-changing devices or actions.\par
\textbf{Iteration.} (1) map Earth to the room and glacier melting to warming; (2) reframe each option by whether it warms or cools the room; (3) keep agreement or ask a heat-focused tiebreaker.\par
\textbf{Return.} The scaffold makes the analogy explicit before extracting the final option.
\end{examplecard}
\end{minipage}\hfill
\begin{minipage}[t]{0.48\linewidth}
\vspace{0pt}
\begin{examplecard}{SuperGPQA}
\small
\textbf{Problem shape.} A chemical-engineering absorption-tower item asking how outlet concentration changes when gas flow doubles.\par
\textbf{Strong-LLM harness.} Derives transfer-unit quantities analytically, computes the new outlet concentration, and asks the solver to verify the closest option.\par
\textbf{Iteration.} (1) compute the old and new transfer-unit values; (2) prompt for independent verification; (3) use a second verification pass; (4) majority-vote or tiebreak if extracted options differ.\par
\textbf{Return.} The scaffold combines deterministic calculation with solver-side option checking for a high-difficulty technical row.
\end{examplecard}
\end{minipage}
\caption{Strong-LLM harness examples for OpenBookQA and SuperGPQA. The first is analogy decomposition; the second is analytic calculation plus verification.}
\label{fig:harness-examples-3}
\end{figure}

\begin{figure}[H]
\centering
\begin{minipage}[t]{0.48\linewidth}
\vspace{0pt}
\begin{examplecard}{Time-MQA}
\small
\textbf{Problem shape.} Classify a short numeric sequence that decreases to a minimum and then increases.\par
\textbf{Strong-LLM harness.} Presents the same signal through trend summaries and compact trend-classification prompts.\par
\textbf{Iteration.} (1) summarize the down-then-up pattern; (2) ask a separate classification prompt; (3) ask a compact tiebreaker distinguishing cyclical from monotonic behavior; (4) vote across extracted answers.\par
\textbf{Return.} The scaffold tests whether repeated views of the same numeric trace stabilize the solver answer.
\end{examplecard}
\end{minipage}\hfill
\begin{minipage}[t]{0.48\linewidth}
\vspace{0pt}
\begin{examplecard}{MedQA}
\small
\textbf{Problem shape.} A medical-board item about suspected digoxin toxicity in an older patient with gastrointestinal symptoms, visual disturbance, bradycardia, renal risk, and hyperkalemia.\par
\textbf{Strong-LLM harness.} Builds a broad clinical analysis, then narrows to digoxin toxicity, compares management options, and asks for a final consensus answer.\par
\textbf{Iteration.} (1) analyze symptoms, medications, potassium, pulse, and renal risk; (2) test the digoxin-toxicity hypothesis; (3) evaluate each answer option; (4) run verification and final consensus prompts.\par
\textbf{Return.} The harness shows why the medical-domain result must be read as benchmark evidence, not clinical deployment evidence.
\end{examplecard}
\end{minipage}
\caption{Strong-LLM harness examples for Time-MQA and MedQA. These rows show repeated trend framing and multi-step clinical verification.}
\label{fig:harness-examples-4}
\end{figure}

\begin{figure}[H]
\centering
\begin{minipage}[t]{0.48\linewidth}
\vspace{0pt}
\begin{examplecard}{PhysicsQA}
\small
\textbf{Problem shape.} Derive the standing-wave amplitude from two equal waves traveling in opposite directions.\par
\textbf{Strong-LLM harness.} Splits the task into algebraic superposition, interpretation of the amplitude coefficient, and final option validation.\par
\textbf{Iteration.} (1) ask for the cosine-sum derivation; (2) ask which amplitude convention applies after deriving $2A\cos(2\pi x/\lambda)$; (3) verify that physical amplitude is non-negative; (4) select the absolute-value option.\par
\textbf{Return.} The scaffold separates symbolic manipulation from the physical interpretation that determines the answer.
\end{examplecard}
\end{minipage}\hfill
\begin{minipage}[t]{0.48\linewidth}
\vspace{0pt}
\begin{examplecard}{Humanity's Last Exam Pilot}
\small
\textbf{Problem shape.} Compare candidate phylogenetic trees by maximum parsimony over alien-species morphological traits.\par
\textbf{Strong-LLM harness.} Extracts variable traits, builds a binary character matrix, asks for matrix verification, evaluates candidate topologies, and then performs a manual-style Fitch parsimony check on promising trees.\par
\textbf{Iteration.} (1) extract traits; (2) verify the matrix; (3) compute parsimony scores across candidates; (4) re-check the closest candidates with explicit per-character changes; (5) return the final tree option.\par
\textbf{Return.} The HLE row is shown as a hard pilot example and remains outside the balanced primary analysis.
\end{examplecard}
\end{minipage}
\caption{Strong-LLM harness examples for PhysicsQA and the HLE pilot. These rows show algebra-then-interpretation and matrix-then-tree-search workflows.}
\label{fig:harness-examples-5}
\end{figure}

\section{Prompt Contract and Answer Channels}
\label{app:prompts}

\begin{promptbox}{Direct Baseline Prompt}
\small
The prompt asks the direct solver to solve a domain-specific MCQA item and output only the option letter. It does not request exposed reasoning. This channel measures conventional direct answer selection under a strict output contract.
\end{promptbox}

\begin{promptbox}{Generator Role}
\small
The prompt frames the generator as a domain expert and Python coder. It tells the generator that the target solver may lack subject knowledge and that additional context or logical flow can help. The generator may split the item into subquestions, create loops, and use repeated solver calls.
\end{promptbox}

\begin{promptbox}{Generated-Code Contract}
\small
The generated program must return the assisted solver answer, the generator-side answer, and a generator-estimated difficulty. It may use solver calls and answer extraction. The prompt states that the answer should not be hard-coded, but the current evaluation treats this as a prompt instruction rather than an enforced guarantee.
\end{promptbox}

\begin{auditbox}{Interpretation Checklist}
\small
\begin{itemize}[leftmargin=*]
  \item Direct and assisted channels use the same target solver within a dataset--solver pair.
  \item Assisted inference has a larger budget than direct inference.
  \item The generator-side answer is diagnostic and should not be collapsed with assisted solver accuracy.
  \item Zero-baseline gains are prompt/scaffold diagnostics, not deployment evidence.
  \item Literal answer assignments and extraction failures are part of the empirical result.
\end{itemize}
\end{auditbox}

\section{Claims and Scope}
\label{app:claims}

\begin{auditbox}{Supported Claim}
\small
In the observed non-zero-baseline partition, CGR is associated with a +28.10 percentage-point macro gain in assisted accuracy over direct accuracy, with a pair-bootstrap interval of [20.32, 36.43]. Under the stricter $A_b>30\%$ gate, the macro gain is +14.11 points.
\end{auditbox}

\begin{auditbox}{Unsupported Claims}
\small
The current evidence does not support claims that CGR is equal-cost, universally beneficial, clinically or operationally safe, or causally isolated from generator knowledge and extra solver calls. It also does not support a claim that the no-hard-coding rule is enforced by runtime validation.
\end{auditbox}

\begin{auditbox}{Future Evaluation Needs}
\small
The strongest next ablations are matched-budget direct self-consistency, repeated generated-program sampling, stricter option validation, sandboxed execution, and source-grounded checks for medicine and industrial root-cause domains.
\end{auditbox}


\end{document}